\documentclass[acmtog,preprint]{acmart}
\renewcommand\footnotetextcopyrightpermission[1]{}

\setcitestyle{nosort,square}

\usepackage{mathrsfs}
\usepackage{tikz}
\usetikzlibrary{arrows.meta}
\usepackage{overpic}
\usepackage{pict2e}
\usepackage{dsfont}
\usepackage[most]{tcolorbox}
\usepackage[outline]{contour}
\usepackage{standalone}
\usepackage[normalem]{ulem}

\usepackage{microtype}      %
\usepackage[capitalize]{cleveref}
\usepackage{xspace}         %
\usepackage{mathtools}      %
\usepackage{nicefrac,xfrac} %
\usepackage{stmaryrd}  %
\usepackage{newfloat}  %
\usepackage{savesym}
\usepackage{algorithm}
\usepackage{algorithmicx}
\usepackage[noend]{algpseudocode} 
\algrenewcommand\textproc{}
\usepackage{graphicx}
\usepackage{grffile}  %
\usepackage{soul}
\usepackage{wrapfig}
\usepackage{bbm}
\usepackage{flushend}
\usepackage{tabularx}
\usepackage{ctable}            %
\usepackage[a]{esvect}           %
\usepackage{multirow}

\usepackage{wrapfig}
\usepackage{textcomp}    %
\usepackage{enumitem}    %

\definecolor{Red}{rgb}{1,0.45,0.45}
\definecolor{Green}{rgb}{0.44,0.76,0.33}
\definecolor{Blue}{rgb}{0.46,0.7,1}
\definecolor{DarkGreen}{rgb}{0.0,0.5,0.0}
\definecolor{DomainColor}{gray}{0}
\definecolor{ExtSpaceColor}{rgb}{0.86,0,0.86}
\definecolor{ExtSpaceColorTwo}{rgb}{0.86,0.5,0.86}
\definecolor{TargetSpaceColor}{hsb}{0.45,0.87,0.81}
\definecolor{TargetMeasureColor}{rgb}{0.3,0.3,0.3}
\definecolor{RedClassColor}{hsb}{0.99,0.62,0.9}
\definecolor{BlueClassColor}{hsb}{0.61,0.6,0.9}
\definecolor{PurpleClassColor}{rgb}{0.86,0.35,0.88}
\definecolor{ProjectionColor}{rgb}{0.95,0.57,0.00}
\newcommand{\commentOutText}[1]{}
\savesymbol{Comment}
\algrenewcommand\algorithmicindent{3mm}
\DeclareFloatingEnvironment[
    name=Alg.,
    placement=tbhp,
    within=none,
]{pseudocode}
\crefname{pseudocode}{Alg.}{Algs.}
\Crefname{pseudocode}{Algorithm}{Algorithms}

\DeclareGraphicsExtensions{.ai,.pdf,.jpg,.png}

\DeclareDocumentCommand{\Outlined}{ O{black} O{white} O{0.55pt} m }{%
    \contourlength{#3}%
    \contour{#2}{\textcolor{#1}{#4}}%
}

\newcommand{\undefinecolor}[1]{\expandafter\let\csname\string\color@#1\endcsname\undefined}

\newcommand{\tsubheader}[1]{%
  \kern -0.5em
  \textit{#1}%
  \kern 0.5em
  \leaders\hbox{%
    \raisebox{.2\ht\strutbox}{\rule{1pt}{.4pt}}%
  }\hfill
  \kern0pt
}

\newcommand{\sdensity}{\bm{s}}
\newcommand{\sprob}{\bm{p}}
\newcommand{\unif}{\bm{u}}

\newcommand{\rnoisy}{\bm{L}^{\text{noisy}}}
\newcommand{\rlayernoisy}{\bm{L}}
\newcommand{\rlayerdenoised}{\hat{\bm{L}}}

\newcommand{\rdenoised}{\bm{L}^{\text{denoised}}}

\newcommand{\kerneltemporal}{\prescript{\text{t}}{}{\kappa}}
\newcommand{\kernelups}{\prescript{\text{u}}{}{\kappa}}
\newcommand{\kerneldenoise}{\prescript{\text{d}}{}{\kappa}}
\newcommand{\rreference}{\bm{L}^{\text{ref}}}
\newcommand{\rrender}{\bm{S}}
\newcommand{\rextra}{\bm{r}}
\newcommand{\rholdout}{\bm{L}^{\text{h}}}
\newcommand{\tonemap}{\tau}
\newcommand{\tdenoised}{\bm{I}^{\text{denoised}}}

\newcommand{\p}{\partial}
\newcommand{\unifdist}{\mathcal{U}}

\newcommand{\E}{\mathop{\mathbb{E}}}
\newcommand{\xy}{x\mkern-1.5mu y}

\usepackage{bm}
\usepackage[most]{tcolorbox}
\tcbset{
  roundedmathbox/.style args={#1,#2}{
    enhanced,                 %
    frame hidden,
    colback=#2!15,          %
    arc=4mm,                  %
    left=2mm, right=2mm,      %
    top=4mm, bottom=2mm,      %
    overlay={
      \node[
        anchor=north west,
        font=\footnotesize\sffamily\itshape,
        text=#2!80!black,
        xshift=3mm,
        yshift=-2mm,
        inner sep=0mm
      ] at (frame.north west) {#1};
    }
  }
}

\begin{document}
\sloppy

\fancypagestyle{standardpagestyle}{%
  \fancyfoot{}%
  \fancyfoot[C]{\small\textsc{Author preprint}}%
}
\pagestyle{standardpagestyle}

\makeatletter
\g@addto@macro\ps@firstpagestyle{%
  \fancyfoot{}%
  \fancyfoot[C]{\small\textsc{Author preprint}}%
}
\makeatother

\title{Forget Superresolution, Sample Adaptively (when Path Tracing)}

\author{Martin B\'alint}
\affiliation{%
 \institution{Max Planck Institute for Informatics}
 \streetaddress{Campus E1 4}
 \city{Saarbr\"ucken}
 \postcode{66123}
 \country{Germany}}
\email{mbalint@mpi-inf.mpg.de}

\author{Corentin Sala\"un}
\affiliation{%
 \institution{Max Planck Institute for Informatics}
 \streetaddress{Campus E1 4}
 \city{Saarbr\"ucken}
 \postcode{66123}
 \country{Germany}}
\email{csalaun@mpi-inf.mpg.de}

\author{Hans-Peter Seidel}
\affiliation{%
 \institution{Max Planck Institute for Informatics}
 \streetaddress{Campus E1 4}
 \city{Saarbr\"ucken}
 \postcode{66123}
 \country{Germany}}
\email{hpseidel@mpi-inf.mpg.de}

\author{Karol Myszkowski}
\affiliation{%
 \institution{Max Planck Institute for Informatics}
 \streetaddress{Campus E1 4}
 \city{Saarbr\"ucken}
 \postcode{66123}
 \country{Germany}}
\email{karol@mpi-inf.mpg.de}

\begin{teaserfigure}
\includegraphics[width=\textwidth]{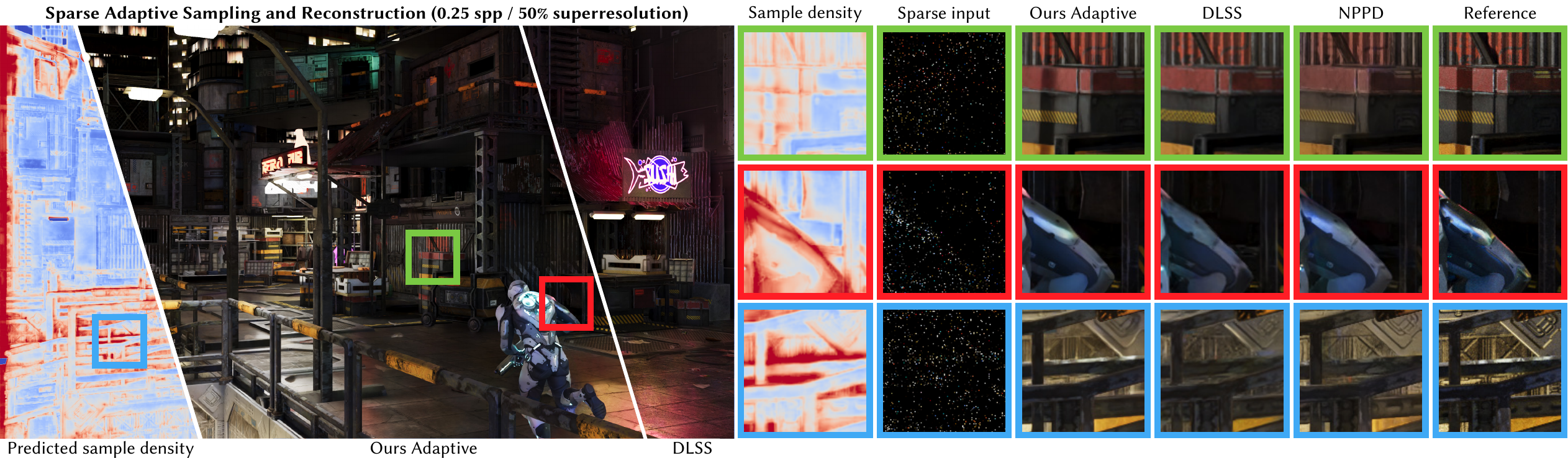}
  \caption{
    \textbf{Sparse adaptive sampling at 0.25~spp.} Our method learns to predict a sampling density map (left, blue means fewer samples, red means more) that concentrates the limited rendering budget on perceptually important regions. Compared to state-of-the-art methods, DLSS 4 Ray Reconstruction \citep{nvidia_dlss4_research} and NPPD \citep{Balint2023Neural}, our method recovers many more details. In the first (green) inset, our method tracks the shadow boundary and achieves a sharp reconstruction, whereas others miss it entirely. In the second (red) inset, we recover the metallic visual of the character's suit, even in motion, whereas others produce a flat look. In the third (blue) inset, our method recovers the specular highlight on a disoccluded piece of railing, whereas others offer little lighting detail due to the complex geometry and camera motion.
}
  \label{fig:Teaser}
\end{teaserfigure}

\begin{abstract}

Real-time path tracing increasingly operates under extremely low sampling budgets, often below one sample per pixel, as rendering complexity, resolution, and frame-rate requirements continue to rise. While super-resolution is widely used in production, it uniformly sacrifices spatial detail and cannot exploit variations in noise, reconstruction difficulty, and perceptual importance across the image. Adaptive sampling offers a compelling alternative, but existing end-to-end approaches rely on approximations that break down in sparse regimes.
We introduce an end-to-end adaptive sampling and denoising pipeline explicitly designed for the sub-1-spp regime. Our method uses a stochastic formulation of sample placement that enables gradient estimation despite discrete sampling decisions, allowing stable training of a neural sampler at low sampling budgets. To better align optimization with human perception, we propose a tonemapping-aware training pipeline that integrates differentiable filmic operators and a state-of-the-art perceptual loss, preventing oversampling of regions with low visual impact.
In addition, we introduce a gather-based pyramidal denoising filter and a learnable generalization of albedo demodulation tailored to sparse sampling. Our results show consistent improvements over uniform sparse sampling, with notably better reconstruction of perceptually critical details such as specular highlights and shadow boundaries, and demonstrate that adaptive sampling remains effective even at minimal budgets.

\end{abstract}



\maketitle

\section{Introduction}

Real-time physically based rendering is fundamentally limited by the number of light paths that can be computed within tight time budgets. At high display resolutions and frame rates, especially on complex scenes, this budget may fall short of even a single sample per pixel (\textit{spp}). Recent hardware advances have made real-time path tracing increasingly practical \citep{ouyang2021restir}, yet per-sample complexity will continue to rise as rendering pipelines adopt neural materials \citep{Zeltner2024Real-timeNeuralAppearanceModels}, sophisticated light transport, and volumetric effects. As a result, sparse sampling budgets are likely here to stay.

Superresolution has become commonplace in production game engines. Rendering at reduced internal resolution and upscaling the result is well-suited to rasterization, which relies on a fixed pixel grid. Optionally, learned denoisers are paired with superresolution and used to help resolve simpler Monte Carlo rendered effects. However, as rendering moves toward full path tracing, spatial adaptation is becoming increasingly important. Monte Carlo noise, denoiser reconstruction quality, and the perceptual visibility of artifacts all vary across the image depending on content, yet superresolution sacrifices spatial detail uniformly.

Luckily, with full path tracing, we are free to place samples anywhere in the image. A jointly trained sampler and denoiser can, in theory, learn to navigate the interdependencies between noise, reconstruction, and perception, concentrating samples where they matter most for the final image. Our goal is to build the first end-to-end adaptive sampling and denoising pipeline for the sub-1-spp regime.

The first challenge is building differentiable sparse adaptive sampling: we need to backpropagate gradients from the rendered image to the sampler network. Unfortunately, choosing which pixels to sample is a discrete operation, making gradient computation non-trivial. Prior work overcomes this issue by relying on approximations or surrogate gradients that hold reasonably well at moderate sample counts but break down as sampling decreases. Such approaches limit effective adaptive sampling to budgets above 2~spp~\citep{hangming2025streaming}. We introduce a formulation that chooses sampled pixels stochastically, letting us derive expected gradients from per-pixel statistics. We efficiently approximate these gradients, enabling end-to-end training of the sampler network. Our approach remains robust with as few as one sample per hundred pixels in the most sparse regions.

Second, effectively redistributing a limited sampling budget requires a carefully designed optimization objective. Minimizing per-pixel error alone is not well aligned with human perception of image distortions and often leads to suboptimal visual results.
Ideally, the method should learn to prioritize reconstruction in the most perceptually critical regions. Unfortunately, loss formulations used by prior works are not well-aligned with perceptual sensitivity, resulting in apparent over- and undersampling artifacts across the image. The issue is amplified in the sub-1-spp regime, where deviations from reference images are necessarily larger, making it harder for losses to reliably judge which distortions are most noticeable. To this end, we propose a tonemapping-aware pipeline that includes the state-of-the-art perceptual loss, MILO~\citep{cogalan2025milo}.

Third, sparse inputs require architectural changes to the denoiser itself. We introduce a gather-based pyramidal filter that remains stable when most pixels lack samples, and a learnable demodulation mechanism that preserves texture detail without relying on renderer-provided albedo decomposition.

Finally, we demonstrate that stable, consistent adaptive sampling strategies can be learned even at minimal  (\cref{fig:Teaser}). Our adaptive sampling provides notably better quality metrics, in addition to resolving small-but-difficult details such as shadow boundaries and specular highlights. We further show how adaptive sampling benefits from the more precise sample placement afforded by higher resolutions, pointing toward a promising direction for real-time path tracing research.

We summarize our contributions as follows:
\begin{itemize}[nosep]
    \item A stochastic formulation for differentiable sparse adaptive sampling, enabling stable end-to-end training at sub-1-spp budgets.
    \item Tonemapper-aware training pipeline with a family of differentiable filmic curves, allowing the use of state-of-the-art perceptual losses.
    \item A denoising filter adapted to sparse renderings.
    \item Dataset and code of our method (released upon acceptance).
\end{itemize}

\section{Related Work}
\label{sec:relatedwork}

\begin{figure*}
    \centering
    \includegraphics[width=\textwidth]{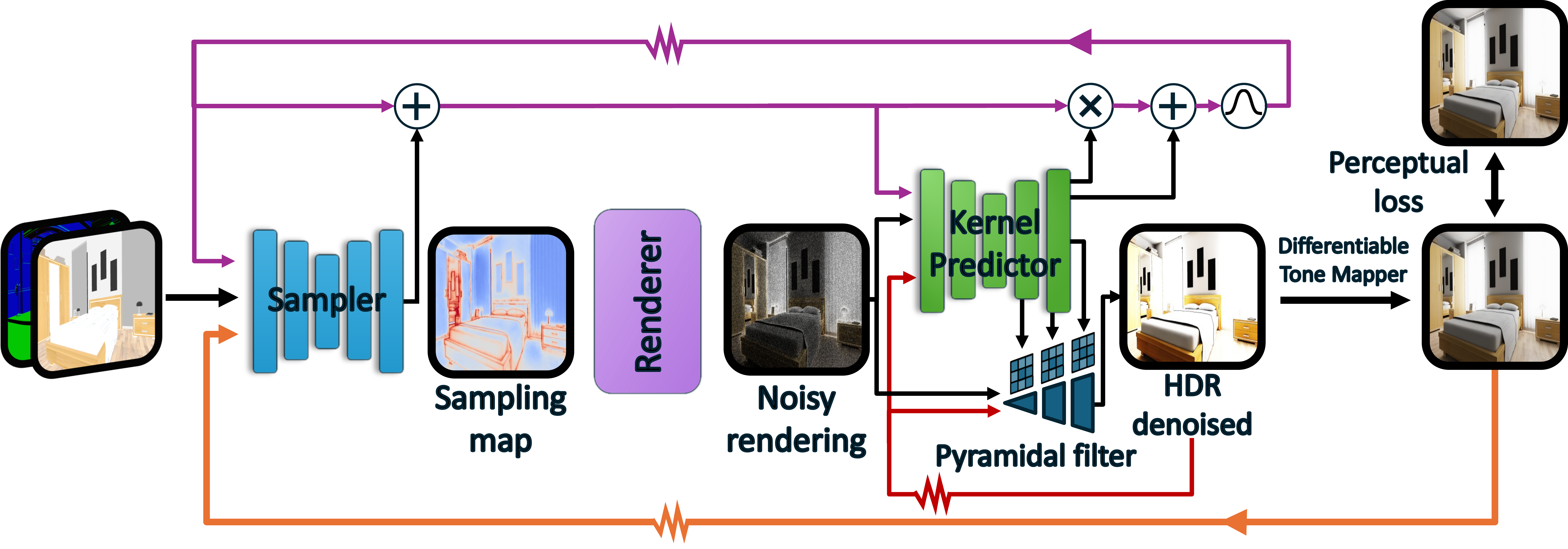}
    \vspace{-0.7cm}
    \caption{\textbf{Overview of our pipeline.} Given GBuffer features and temporal data from the previous frame, the sampler network predicts a continuous sample density map. After stochastic discretization (\cref{sec:sampling}), the renderer traces the requested samples, producing a sparse, noisy image. The denoiser network, which shares latent features with the sampler, predicts weights for our pyramidal gather filter (\cref{sec:implementation}). The denoised HDR output is tonemapped for display and fed back to the sampler in the next frame, closing our perceptual feedback loop (\cref{sec:tonemapper}). }%
    \label{fig:pipeline}
\end{figure*}

\paragraph{Adaptive sampling}

Adaptive sampling plays a crucial role in Monte Carlo rendering, where noise in the produced images is not uniformly distributed across the entire image. Samples from some pixels may exhibit larger variance than others, meaning the number of samples required to achieve satisfactory quality across the image may vary  \cite{mitchell1987generation}. Adaptive sampling addresses this challenge by adjusting the sample count for each pixel to equalise the per-pixel error, ultimately leading to an overall higher image quality within the same rendering time. 

We sort adaptive sampling techniques into two main categories: iterative and prediction-based sampling. In iterative methods, the image is rendered in multiple passes, typically starting with a uniform pass and then updating the sample distribution in subsequent passes based on gathered statistics \cite{bolin1998perceptual,Firmino2023DenoisingAware}. On the other hand, prediction-based sampling utilises guiding information available prior to rendering to predict the sample budget for each pixel in one pass. These methods often rely on neural networks for their robust predictive capabilities. In this work, we focus on this second class of adaptive sampling.

When using both neural-based sampling and denoising, optimising each component separately using simple heuristics only achieves results with limited quality. Therefore, adapting the sampling network to a given denoiser is crucial for achieving high-quality results with fewer samples. For instance, \citet{Vogels2018Denoising} proposed training an error predictor network to estimate regions with higher error and then using it as a sampling map for adaptive sampling. \Citet{Kuznetsov2018Deep} suggested directly utilising backpropagation of the denoising error through the rendering process using a finite difference gradient approximation of the rounded sample count for low sample count adaptive sampling. This method was later extended with temporal processing by \citet{hasselgren2020ntasd}, guiding sampling using the previous denoised frame, further reducing rendering costs. \citet{Salehi2022DeepAdaptive} proposed removing the renderer from the training process by generating path tracing samples from approximate analytic distributions. This method relies on the convergence of the Monte Carlo estimator to a Gaussian distribution as the sample count increases, thanks to the Central Limit Theorem. Unfortunately, Monte Carlo estimators do not follow such an analytic distribution at low sample counts. The closest method to our work from \citet{scardigli2023rl} uses reinforcement learning to optimise the sampling network through a critic network trained to predict the denoiser's expected loss for a given sampling map.

\paragraph{Denoising}

Denoising is a crucial step in rendering pipelines utilising Monte Carlo-based light transport algorithms. Its main goal is eliminating the noise inherent in Monte Carlo processes, allowing for the generation of high-quality images even with limited sampling budgets. Early denoising methods involved local blurring of noisy input images with non-linear filters \cite{Mitchell1987Generating,Lee1990ANote,Rushmeier1994Energy,Xu2005ANovel,Overbeck2009Adaptive}. Subsequent advancements incorporated auxiliary features such as surface normals, albedo, depth, and estimated variance to adapt filters locally, producing more accurate images \cite{McCool1999Anisotropic,Li2012SURE,Rousselle2013Robust,Bitterli2016Nonlinearly}. Recently, the advent of deep learning and corresponding hardware development has spurred the creation of deep learning-based denoisers, leading to significant improvements in denoising quality. These approaches range from trainable models for directly predicting the output image from noisy input \cite{Xiao2020Neural,zhang2024subpixel} to kernel prediction techniques that learn per-pixel kernel parameters \cite{Gharbi2019SampleBased,hofmann2023volume,zhang2024deepzdenoising,lee2024fusion}. Most recently, multiscale kernel prediction and filtering have shown significant improvements in interactive denoising \cite{Balint2023Neural,icsik2021interactive}.

Most denoising methods assume a uniform distribution of samples across the screen space, often assuming samples in every image pixel. However, in the context of adaptive sampling and real-time rendering, it may not always be feasible or desirable to adhere to such constraints. Using a denoiser optimized for a uniform sample distribution in adaptive sampling methods can result in poor outcomes, especially with deep learning methods. Consequently, there has been an exploration into developing denoisers that are aware of adaptive sampling \cite{Salehi2022DeepAdaptive,Firmino2023DenoisingAware}. %

\paragraph{Super-resolution}

Super-resolution is an alternative method to improve performance by rendering at a reduced resolution and upscaling the result to full resolution. This guarantees that all pixels can be sampled, but on a coarser grid. The approach is widely used in real-time pipelines, notably in production methods such as DLSS, FSR, and XeSS. Early learning-based methods include the three-layer CNN of \citet{Ji2016Image} and the Laplacian pyramid network of \citet{Lai2017Deep}. Later work incorporates temporal information and auxiliary buffers for real-time use \citep{Lei2020Neural, zhong2023fusesr}, while generative methods attempt to hallucinate high-frequency details beyond the input signal \citep{Ge2018Image}.

Rendering at reduced resolution followed by super-resolution offers a simple alternative to adaptive sampling, but it lacks spatial selectivity. All regions receive the same effective sampling rate, regardless of their rendering complexity. In contrast to adaptive sampling, this prevents oversampling difficult regions such as reflections, shadows, or high-frequency lighting effects while undersampling simpler areas. In this work, we show that super-resolution may be better suited to rasterization pipelines with uniform per-pixel cost than to Monte Carlo path tracing, particularly in scenarios where noise and information content vary significantly across the image, and efficient reconstruction at low sampling budgets relies on highly targeted sampling strategies.

Super-resolution has also been successfully combined with denoising for path-traced rendering \citep{Kazmierczyk2025joint,Thomas2022Temporally}. The additional denoising stage enables more faithful reconstruction of fine details than super-resolution alone, particularly in regions affected by Monte Carlo noise.

\paragraph{Tone Mapping for neural training}Most learning-based denoising and adaptive sampling methods optimize losses defined in linear HDR space, typically using relative-error metrics such as SMAPE or logarithmic $L1$ losses \cite{Vogels2018Denoising,Kuznetsov2018Deep,Salehi2022DeepAdaptive,Balint2023Neural}. These losses do not account for perceptual adaptation or downstream tonemapping and therefore misrepresent which errors are actually visible. As a result, computation is often wasted in very dark or very bright regions where large linear errors are perceptually negligible. Some approaches instead train directly in tonemapped (SDR) space \cite{hasselgren2020ntasd,OpenImageDenoise,hangming2025streaming}, improving perceptual alignment but introducing bias in kernel-based denoisers and coupling the model to a fixed tonemapping operator. These limitations highlight the need for training strategies that explicitly account for tonemapping and perception when jointly optimizing sampling and denoising.
To address these limitations, we propose an expressive family of parametric, differentiable tone mapping operators suitable for end-to-end learning of adaptive sampling and denoising. In addition, we employ a perceptual error metric as the loss function to guide sample allocation toward image regions containing highly visible and objectionable distortions, as perceived by a human observer.

\section{Overview}

Before detailing our contributions, we outline the pipeline that integrates them. Adaptive sampling and denoising pipelines share a common structure (\cref{fig:pipeline}): a sampler network predicts per-pixel sample densities, the renderer traces paths accordingly, and a denoiser reconstructs the final image from the noisy result. Our pipeline follows this established design, with modifications that enable stable training at sub-1-spp budgets.

The sampler takes GBuffer features---depth, normals, and albedo---alongside the reprojected output and latent features from the previous frame (Orange and Purple loops). From these inputs, it predicts a continuous sample density map that distributes the frame's sampling budget across pixels. This density must then be discretized into integer sample counts before rendering. As we detail in \cref{sec:sampling}, this discretization step is the key obstacle to end-to-end training: deterministic rounding \citep{Kuznetsov2018Deep} produces zero gradients almost everywhere, while existing surrogate approximations become unstable at low sample counts. We propose a stochastic formulation that interprets fractional densities as probabilities, enabling robust gradient estimation even at extreme sparsity.

We obtain the sample density map $\sdensity$ by first applying softmax normalization to one of the sampler network's output channels. We then allocate $\nicefrac{1}{8}$ of the total budget uniformly to ensure numerical stability, and distribute $\nicefrac{7}{8}$ adaptively based on the network's normalized output. We find that distributing a larger fraction of the budget adaptively slows training convergence without yielding any measurable improvement in quality.

After rendering, the denoiser receives the noisy image together with the same auxiliary features and reprojected data from the previous frame (Purple loop) and the previous frame in HDR (Red feedback loop). Rather than predicting pixel values directly, it outputs weights for a pyramidal gather filter that reconstructs the final HDR image. The sampler and denoiser share a latent feature space (Purple feedback loop), enabling them to collaborate: the sampler can learn to place samples where the denoiser struggles, while the denoiser can adapt its reconstruction to the sampler's allocation patterns.

The reconstructed image is then tonemapped for display. We feed the tonemapped output back to the sampler in the next frame (Orange feedback loop), allowing for alignment with perceptual importance. During training, our differentiable parametric tonemapper (\cref{sec:tonemapper}) enables gradient flow through this loop. At inference, any tonemapper can be substituted: the graphics engine simply feeds its output back to the sampler, making integration straightforward. We describe our gathering pyramidal filter and other architectural choices in \cref{sec:implementation}.

Note that, like previous end-to-end approaches, our method does not require a differentiable renderer. Following prior work by \citet{Kuznetsov2018Deep}, we pre-render images at powers-of-two sample counts, which can be combined to produce arbitrary budgets during training.

\section{Differentiable sparse adaptive sampling}
\label{sec:sampling}

Path tracing is inherently discrete: for each pixel, a sample is either computed or not. Continuous sample densities predicted by the sampler network must therefore be converted to integer counts before rendering. Prior work relies on deterministic rounding with surrogate gradient approximations, but this approach fundamentally limits what adaptive sampling can achieve.

We reframe sample allocation as an exploration-exploitation problem. At each pixel, the sampler must balance \emph{exploiting} regions where it confidently knows samples are needed against \emph{exploring} regions where it is uncertain. A well-tuned strategy samples uncertain regions often enough to gather information while concentrating resources where the payoff is known to be high. Deterministic rounding cannot capture this balance: densities of 0.3 spp and 0.0 spp both yield zero samples, even though the former suggests meaningful uncertainty. We propose stochastic rounding, which interprets the sample density as a probability. A density of 0.3 spp corresponds to a 30\% chance of rendering a sample, allowing the network to express confidence continuously and learn to balance exploration and exploitation through gradient-based optimization.

We derive the expected gradient analytically and introduce a relaxed approximation that enables efficient backpropagation while preserving the estimator's statistical properties. Our approach remains robust at densities as low as 0.01 samples per pixel.

\subsection{Deterministic rounding}
\label{sec:det_rounding}

Deterministic rounding for learnable adaptive sampling was first proposed by \citet{Kuznetsov2018Deep}:
\begin{equation}
    \label{eq:kuz_fw}
    \rnoisy_i =
    \begin{cases}
        \frac{\sum_{j=1}^{\text{round}(\sdensity_i)}\rrender_{i,j}}{\text{round}(\sdensity_i)} & \text{if } \sdensity_i \geq 0.5 \\
        0 & \text{otherwise,}
    \end{cases}
\end{equation}
where, for each pixel $i$, the sample density $\sdensity_i$ is rounded to the nearest integer and that many samples $\rrender_{i,j}$ are averaged to produce the noisy estimate $\rnoisy_i$.

The gradient of \cref{eq:kuz_fw} is zero almost everywhere, reducing to Dirac delta impulses at the transition points. This provides no useful signal for optimization. Instead, \citet{Kuznetsov2018Deep} proposed to use a surrogate gradient:
\begin{equation}
    \label{eq:kuz_grad}
    \frac{\p \rnoisy_i}{\p \sdensity_i} =
    \begin{cases}
        \frac{\rreference_i - \rnoisy_i}{\text{round}(\sdensity_i)} & \text{if } \sdensity_i \geq 0.5 \\
        \rreference_i & \text{otherwise.}
    \end{cases}
\end{equation}
Although many later works \citep{hangming2025streaming, Salehi2022DeepAdaptive, hasselgren2020ntasd} adopt this formulation, the fixed threshold at $\sdensity_i = 0.5$ leads to severe failures in sparse regimes. When a spatially extended region is deemed easy to render and $\sdensity_i < 0.5$ for all its pixels, deterministic rounding assigns zero samples to the entire region. This leaves large areas with no illumination information, making accurate reconstruction impossible. In principle, even a few samples distributed across the region would suffice to recover a plausible signal, but the bias of the surrogate gradient in \cref{eq:kuz_grad} seemingly precludes such partial coverage.

Some approaches mitigate this by imposing mandatory samples in regular patterns \citep{hangming2025streaming}. While this prevents the most important artifacts, it does not allow the sampler to express fine-grained confidence or receive meaningful gradients. Thus, the learned allocation strategies remain limited.

\begin{figure}
    \centering
    \includegraphics[width=0.45\columnwidth]{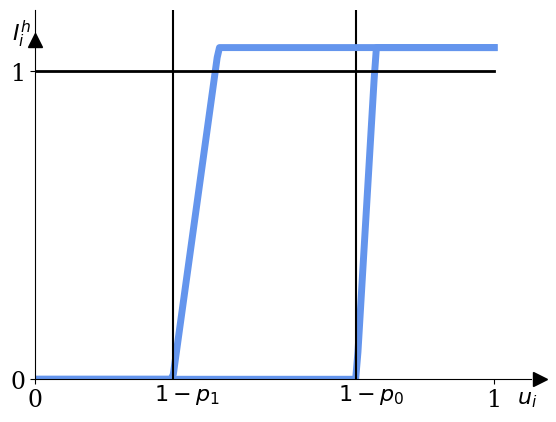}
    \includegraphics[width=0.45\columnwidth]{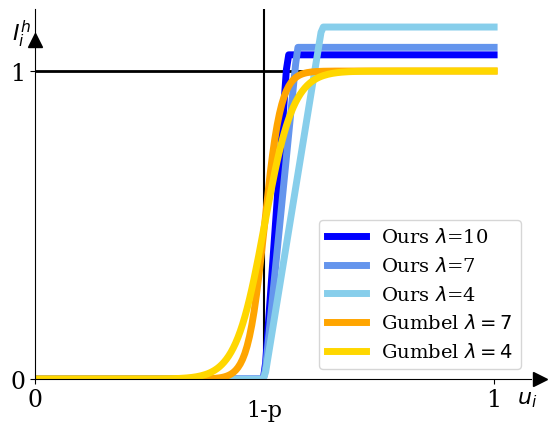}
    \vspace{-0.3cm}
    \caption{
\textbf{Visualization of our relaxed estimator.} On the left, we illustrate the effect of the sampling probability $p$ on the slope of the function at a given temperature parameter $\lambda$. On the right, we compare the estimator at a given value of $p$ for different values of $\lambda$, along with the Gumbel–Softmax estimator for reference.}
\label{fig:Stochastic_rounding_relax}
\end{figure}

\subsection{Stochastic rounding}
\label{sec:stoch_rounding}

We apply the principle of stochastic rounding to the adaptive sampling setting, deriving a formulation that maintains unbiased radiance estimation while enabling gradient-based optimization.

For each pixel $i$, we always allocate the integer part $\lfloor \sdensity_i \rfloor$ samples. Let $\sprob_i = \sdensity_i - \lfloor \sdensity_i \rfloor$ denote the fractional part. We draw a uniform variate $\unif_i \sim \unifdist(0,1)$ and allocate one extra sample whenever $\unif_i < \sprob_i$. We denote this potential extra sample $\rextra_i$ and define its contribution as:
\begin{equation}
\label{eq:holdout}
    \rholdout_i = \rextra_i \,\bm{[}\unif_i < \sprob_i\bm{]}\,.
\end{equation}
The final noisy estimate is the normalized sum of all allocated samples:
\begin{equation}
\label{eq:stoch_rnoisy}
    \rnoisy_i = \frac{\sum_{j=1}^{\lfloor \sdensity_i \rfloor}\rrender_{i,j} + \rholdout_i}{\sdensity_i}\,.
\end{equation}
Importantly, we normalize by the sample density $\sdensity_i$ rather than the discrete number of samples computed. This offers several advantages; first, differentiating this step is trivial for $\sdensity_i > 0$. Second, $\E[\rnoisy_i] = \E[\rrender_i]$, by this construction, making the estimator unbiased across the entire density range. Finally, normalization by $\sdensity_i$ naturally upweights pixels with the extra sample taken, providing inverse-variance weighting. Intuitively, this formulation remains well-defined even when $\sdensity_i \ll 1$: pixels with low densities simply have a correspondingly low probability of being sampled, but when they are, their contribution is appropriately scaled.

Deriving gradients through stochastic processes is a well-established technique for differentiating discrete operations \citep{li2018differentiable}. The expected gradient of the loss with respect to $\sdensity_i$ exists and is well-behaved. However, computing this gradient exactly requires edge sampling of each discontinuity, effectively running the denoiser for each pixel, which is prohibitively expensive for training. We therefore introduce a relaxed formulation that enables efficient gradient computation via standard backpropagation.

\subsection{Relaxed stochastic rounding}
\label{sec:relaxed_stoch_rounding}

\begin{figure}
    \centering
    \begin{tabular}{ccc}
        \includegraphics[width=0.3\columnwidth]{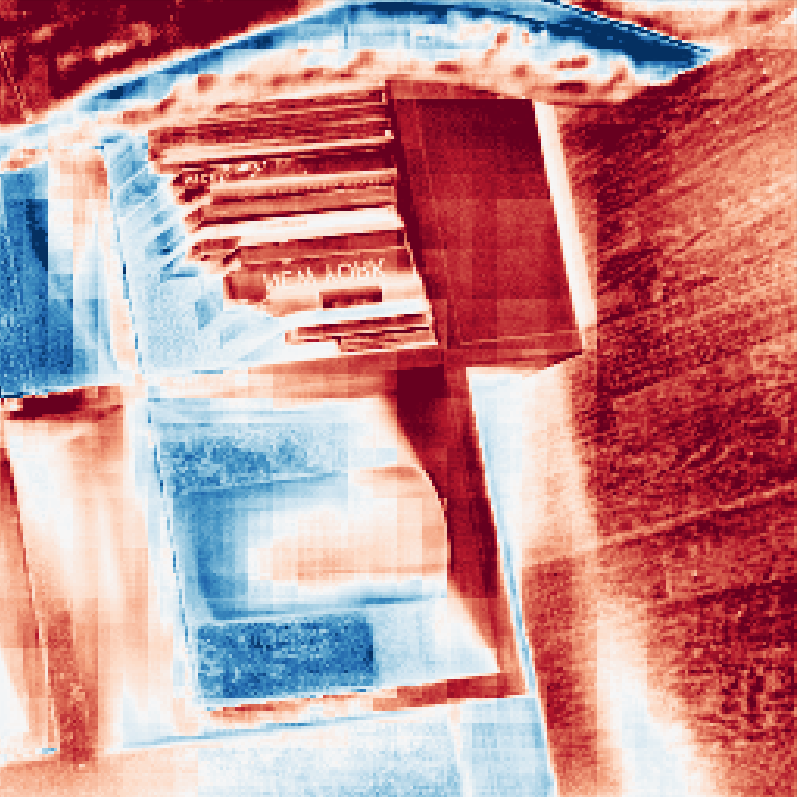} & \includegraphics[width=0.3\columnwidth]{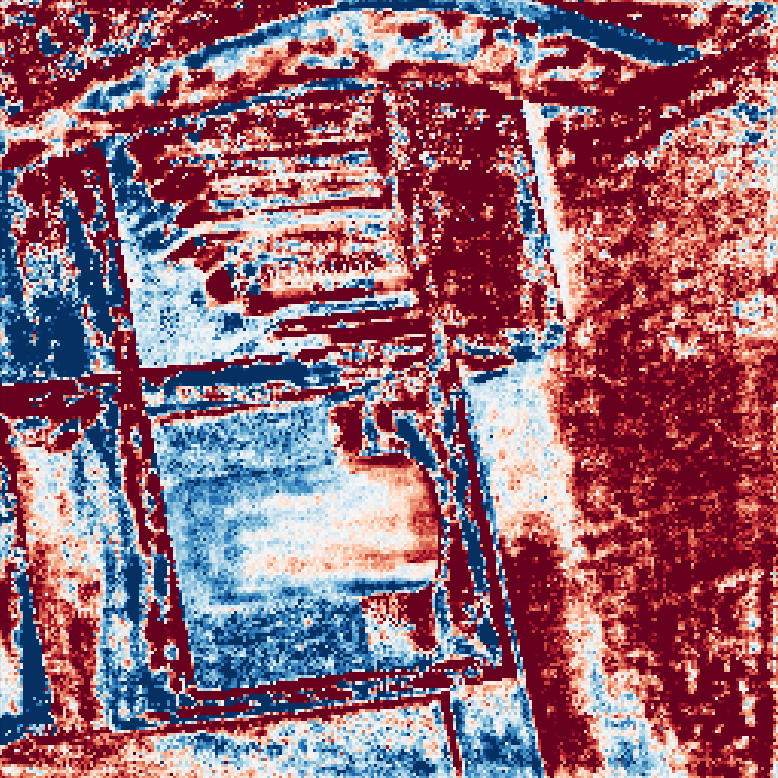} & \includegraphics[width=0.3\columnwidth]{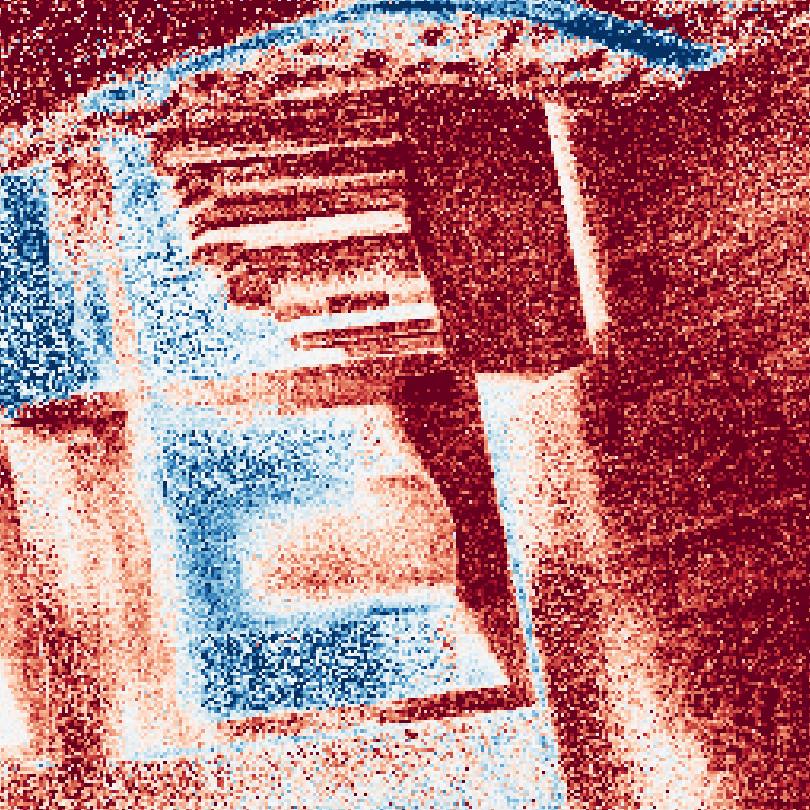}  \\
        Finite difference & Straight through & Ours\\
        gradient & gradient &
    \end{tabular}
    \vspace{-0.4cm}%
    \caption{\textbf{Gradient comparison}. We compare the straight-through estimator and our proposed relaxed estimator to the ground truth gradient estimated via finite differences. Our estimator is closest to the finite-difference reference, leading to more stable optimization due to lower gradient bias.}
    \label{fig:gradient_comparison}
\end{figure}

Following the Gumbel–Softmax relaxation \citep{jang2017categoricalreparameterizationgumbelsoftmax, maddison2017concretedistributioncontinuousrelaxation}, we replace the discontinuous step function in \cref{eq:holdout} with a smooth approximation. While effective for gradient propagation, this relaxation is ill-suited for adaptive Monte Carlo sampling. The resulting sigmoid-like transition has non-zero tails, allowing bright samples to ``shine through'' in regions that should receive none. Although these tails narrow with increasing temperature parameters, rare outlier draws still occur, resulting in test–train mismatch.

We design our relaxation to avoid this issue. Rather than a symmetric sigmoid, we use a clipped linear ramp whose base sits exactly at the sampling threshold:
\begin{equation}
    \label{eq:relaxed_holdout}
    \rholdout_i
    = \rextra_i \cdot h \cdot
    \min\Bigl(
        \max\bigl(
            \tfrac{\lambda}{\sprob_i}
            (\unif_i + \sprob_i - 1),
        0 \bigr), 1
    \Bigr) \text{,}
\end{equation}
where the temperature $\lambda \geq 1$ controls the slope of the ramp. With probability $1 - \sprob_i$, the uniform variate $\unif_i$ falls below the ramp entirely, yielding $\rholdout_i = 0$ with zero gradient. In this case, no sample needs to be computed. Only when $\unif_i$ falls within the linear region (probability $\sprob_i / \lambda$) or above (probability $\sprob_i (1 - 1/\lambda)$) we allocate a sample. This behavior can be replicated exactly at test time: we simply skip sample computation when $\unif_i < 1 - \sprob_i$. \Cref{fig:Stochastic_rounding_relax} depicts our relaxed stochastic rounding function.

As $\lambda \to \infty$, the ramp converges to the hard step in \cref{eq:holdout}. For finite $\lambda$, the ramp produces nonzero gradients over the linear region, proportional to $\sprob_i / \lambda$. Because the ramp is asymmetric, the expected contribution of the holdout sample would be attenuated without correction. We compensate with a factor:
\begin{equation}
    \label{eq:h_factor}
    h = \frac{2\lambda}{2\lambda - 1}\,,
\end{equation}
which rescales the ramp such that $\E[\rholdout_i]$ matches the variant in \cref{eq:holdout}, preserving $\E[\rnoisy_i] = \E[\rrender_i]$. Geometrically, we preserve the area under the curve as seen in \cref{fig:Stochastic_rounding_relax}.

The choice of $\lambda$ presents a tradeoff: larger values produce gradients closer to the true expected gradient but with higher variance. Unlike Gumbel-Softmax, which requires a high temperature to avoid significant train-test mismatch, our formulation is robust across a wide range of temperatures. Lower values of $\lambda$ widen the linear region, adding noise to $\rholdout_i$ when $\unif_i$ falls within it, but this effect quickly diminishes with increasing $\lambda$. Consequently, we observe that ablations trained with $\lambda = 8$, $10$, and $12$ all fall within run-to-run variance. Gradient norms remain comparable to the uniform sampling baseline, indicating that these $\lambda$ values are not large enough to affect training stability either. In practice, we use $\lambda = 10$. We further compare our relaxed gradient estimator to the straight-through estimator \citep{bengio2013estimating} in \cref{fig:gradient_comparison}.

\section{Tonemapper aware training}
\label{sec:tonemapper}

Adaptive sampling trained end-to-end relies on the loss to guide sample allocation. Under low sample budgets, rendering distortions can be substantial. In this regime, a perceptual loss is needed to weigh distortions against each other and prioritize sampling of those that are more visible and objectionable. Optimizing strict per-pixel fidelity is therefore less effective than minimizing perceptual artifact visibility, which can be further altered by downstream display-referred tone mapping \citep{reinhard2010high}. Ignoring these factors can lead to oversampling regions where errors are difficult to perceive, while undersampling regions where additional samples would yield visible improvements.

We address these issues with three modifications to existing approaches. First, we apply a family of differentiable tone mapping operators (TMO) to both network outputs and reference images. Differentiability ensures that gradients can flow backward during training, while at inference, any tonemapper can be used. Second, we provide the sampler network with the previous frame’s tonemapped output. This enables the sampler to automatically adapt to the tonemapped appearance and integrates trivially into existing rendering pipelines for inference. Finally, instead of computing losses in linear or logarithmic color spaces, we evaluate errors in the tone-mapped domain using the state-of-the-art perceptual loss MILO \citep{cogalan2025milo}. This leverages the fact that such perceptual metrics are trained and applied in the sRGB domain, where large-scale human-labeled image quality datasets are predominantly available.

\begin{figure}
    \centering
    \includegraphics[width=0.9\columnwidth]{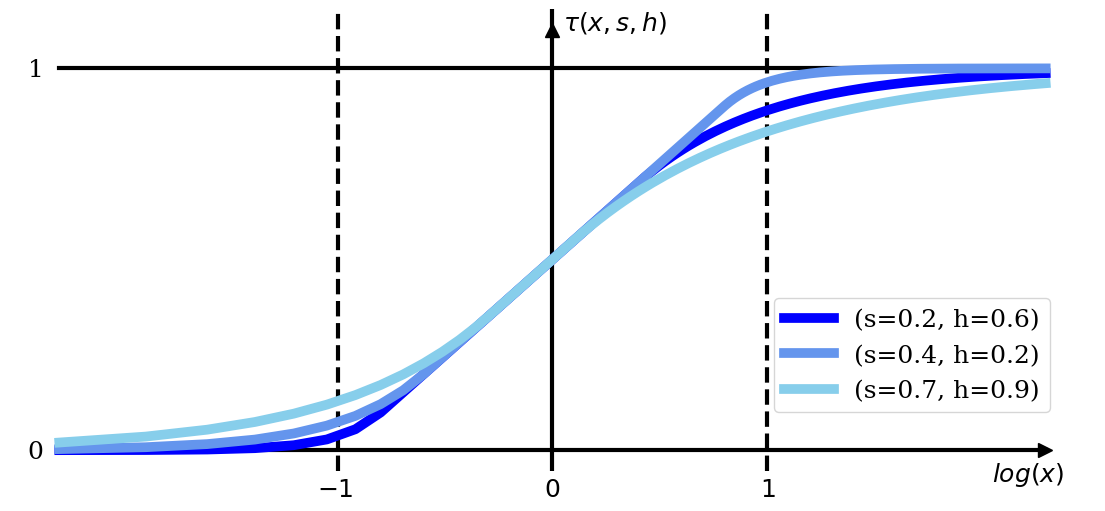}
    \vspace{-0.4cm}
    \caption{\textbf{Visualization of our differentiable tone-mapping operator (TMO) for three parameter settings.} The parameters $s$ and $h$ control the compression of the lower (toe) and upper (shoulder) luminance tails, respectively. The curves are plotted as a function of the logarithm of radiance.}
    \label{fig:diff_TMO}
\end{figure}

\paragraph{Differentiable Parametric Filmic Tonemapping.} Generalization at test time demands training with a diverse set of differentiable TMOs. First, we transform the output of our pyramidal denoising filter to the logarithmic colorspace and apply basic augmentations:
\begin{itemize}
    \item \textbf{Exposure} $k$: additive offset in log-space shifts mid-gray.
    \item \textbf{Contrast} $\alpha$: multiplicative scale about $0$ adjusts curve steepness.
    \item \textbf{Saturation} $\beta$: per-channel scaling around the log-RGB mean modulates color vividness.
\end{itemize}
More formally,
\begin{equation}
    \tdenoised = \textrm{sRGB}(\tonemap(\alpha(\log \rdenoised \odot \beta + k))) \text{,}
\end{equation}
where $\odot$ is a shorthand for the saturation adjustment, and $\tonemap$ is our differentiable filmic curve with further shadow $s\in(0,1)$ and highlight $h\in(0,1)$ compression controls:
\begin{equation}
    \tonemap(x;\,s,h) =
    \begin{cases}
        \frac{s}{2}\exp\left(\frac{x + 1 - s}{s}\right)
        & \text{if } x < s - 1 \\
        \frac{1 + x}{2}
        & \text{if } s - 1 \le x < 1 - h \\
        1 -\frac{h}{2}\exp\left(\frac{-(x + h - 1)}{h}\right)
        & \text{if } x \ge 1 - h \text{.}
    \end{cases}
\end{equation}
When $s,h \to 1$, $\tonemap$ approximates a sigmoid (equivalent to Photographic TMO in log-space \cite{Reinhard2002Photographic}). When $s$ or $h \to 0$, it produces a sharp toe and shoulder. This two-parameter function shape is analogous to common filmic operators \citep{hable2010hdr,Hable2017Filmic, SMPTE2065-1}. \Cref{fig:diff_TMO} shows the tone mapping operator for 3 different sets of parameters $s$ and $h$.

This tone mapper is well-suited for gradient-based training. If we omit the constant factor from the contrast and saturation adjustments for brevity, it is possible to bound the derivatives of our TMO as:

\noindent
\begin{subequations}
\begin{minipage}{0.49\linewidth}
    \begin{equation}
        0 < \frac{\p \tonemap}{\p \log \rdenoised_{t}} < \frac{1}{2} \text{,}
    \end{equation}
\end{minipage}
\hfill
\begin{minipage}{0.49\linewidth}
    \begin{equation}
        0 < \frac{\p \tonemap}{\p \rdenoised_{t}} < \frac{e}{2} \text{.}
    \end{equation}
\end{minipage}
\end{subequations}
\vspace{1.2mm}

\noindent Such bounded derivatives avoid gradient explosions or vanishing, ensuring robust training.

\paragraph{Discussion}
Although we do not claim state-of-the-art perceptual quality, our family of TMO is comparable to widely used operators and suffices as a robust training surrogate. %
Importantly, the tone mapping used at test time can differ from the differentiable TMOs employed during training. The sampler learns to respond to tone-mapped appearance rather than to a specific operator, and the proposed TMOs are sufficiently expressive to generalize across different tone-mapping choices. The inference-time TMO does not need to be differentiable, as we do not rely on inference-time backpropagation. For HDR-only workflows, one could instead use the display transfer function.

\paragraph{Perception-based loss}
Per-pixel error metrics such as L1, L2, and contrast-normalized variants including the symmetric mean absolute percentage error (SMAPE) are commonly used in image reconstruction tasks such as denoising and super-resolution (\cref{sec:relatedwork}). However, when employed to guide adaptive sampling, such losses can be suboptimal, as they assign equal importance to all errors regardless of their visual impact. In dark image regions, Weber-like behavior breaks down, and visual sensitivity is instead governed by the De Vries–Rose law \cite{vanNes1967CSF}, causing these losses to severely overestimate perceptual error and thus oversample such regions. This effect is particularly pronounced for SMAPE, whose Michelson-contrast–like formulation yields small denominators and can indicate near-maximal errors even for barely visible distortions. Conversely, at high luminance levels, all these losses ignore HVS gain control and compressive response \cite{shapley1984visual}, again inflating reported errors. While logarithmic scaling can mitigate oversampling at high luminance, it further exacerbates excessive sample allocation in dark regions. Moreover, per-pixel losses neglect contrast masking and spatial pooling effects, whereby errors become less visible in textured regions \cite{legge1980vision,foley1994human}. In particular, visual masking can permit substantial reductions in required sample density in the presence of strong contrast patterns \cite{myszkowski1998visibility,ramasubramanian1999multiscale}. These shortcomings may be further aggravated by tone mapping, which can significantly alter local luminance and contrast relationships, further decoupling loss predictions from perceived error.

To address these limitations, we instead employ perceptual losses computed on tone-mapped pixel values. Specifically, we adopt the MILO metric \citep{cogalan2025milo}, which implicitly captures %
masking effects and models distortion visibility across both dark and bright regions. MILO is trained on distortion types relevant to our setting, including residual noise from denoising, over-blurring due to undersampling, and hue shifts similar to those introduced by tone mapping. As MILO is trained on human mean opinion score (MOS) data, it weights errors according to their perceived visibility and annoyance, while downweighting distortions below perceptual thresholds. This task-informed training places MILO closer to perception-based error visibility prioritization, in contrast to SSIM/MS-SSIM \cite{wang2004image,Wang2003d}, which emphasizes structural fidelity, and LPIPS \cite{zhang2018lpips}, which focuses on visual similarity. %
Finally, from a practical standpoint, the visibility maps produced by MILO are more efficient to backpropagate than LPIPS feature activations, resulting in more efficient training. Moreover, prior applications of MILO as a loss function in image restoration tasks demonstrate its ability to reconstruct high-frequency content, which is desirable in our setting.

\section{Implementation}
\label{sec:implementation}

\subsection{Gathering Pyramidal Filter}
\label{sec:kernels}

\begin{figure}
    \centering
    \begin{tabular}{c|c}
        \textbf{Gathering} & \textbf{Scattering}\\
        \includegraphics[width=0.45\columnwidth]{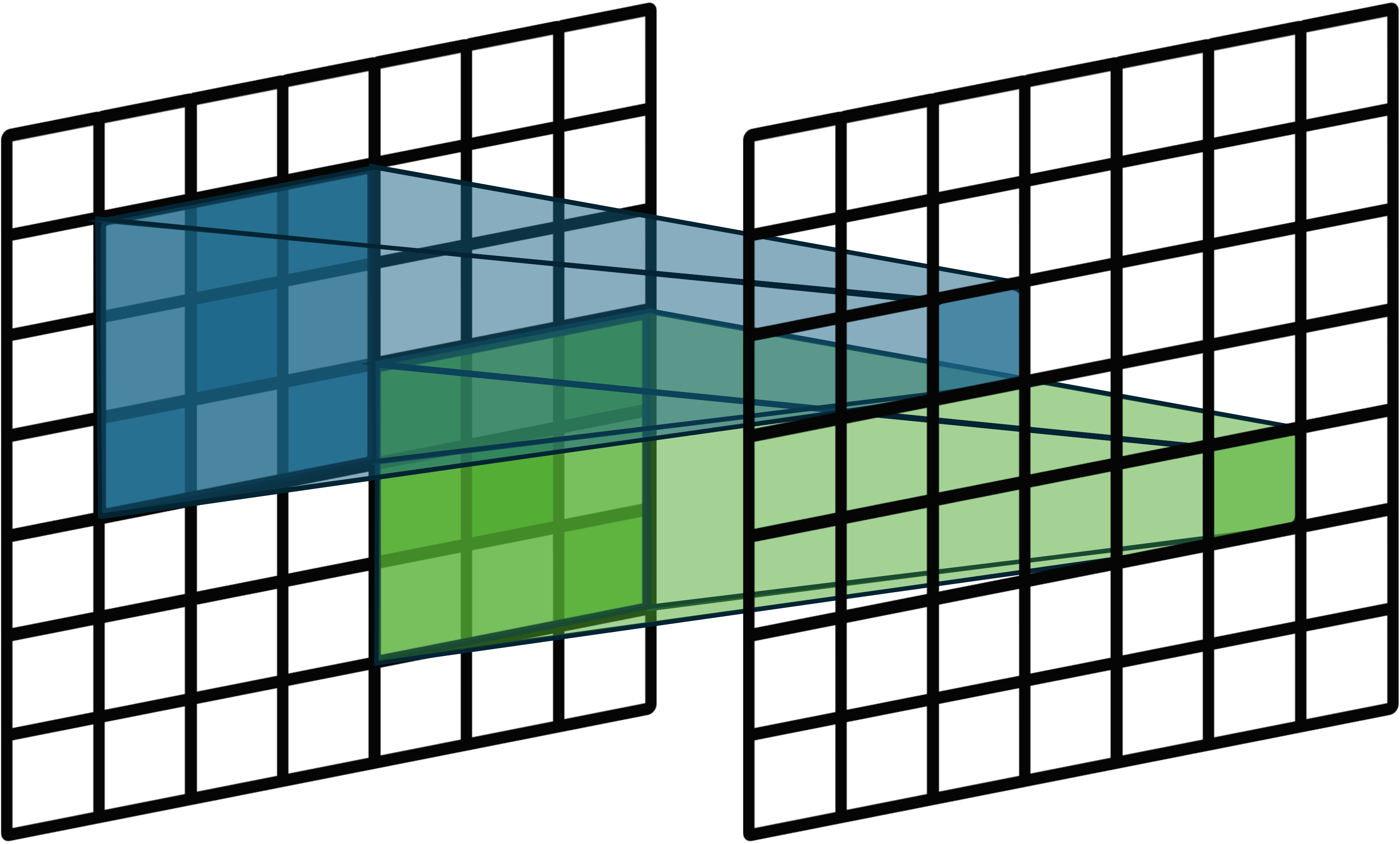} & \includegraphics[width=0.45\columnwidth]{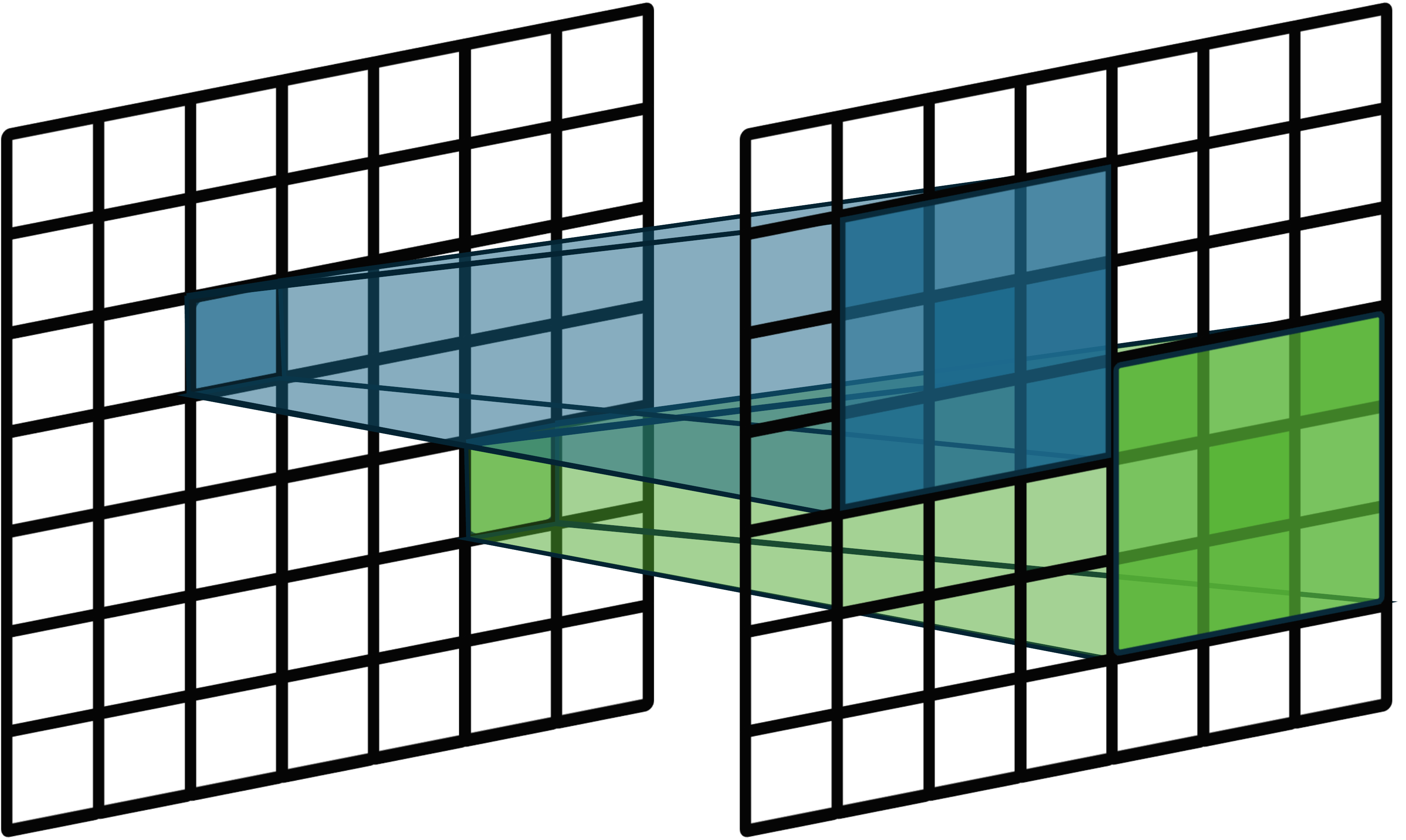} \\[2mm]
        \includegraphics[width=0.35\columnwidth]{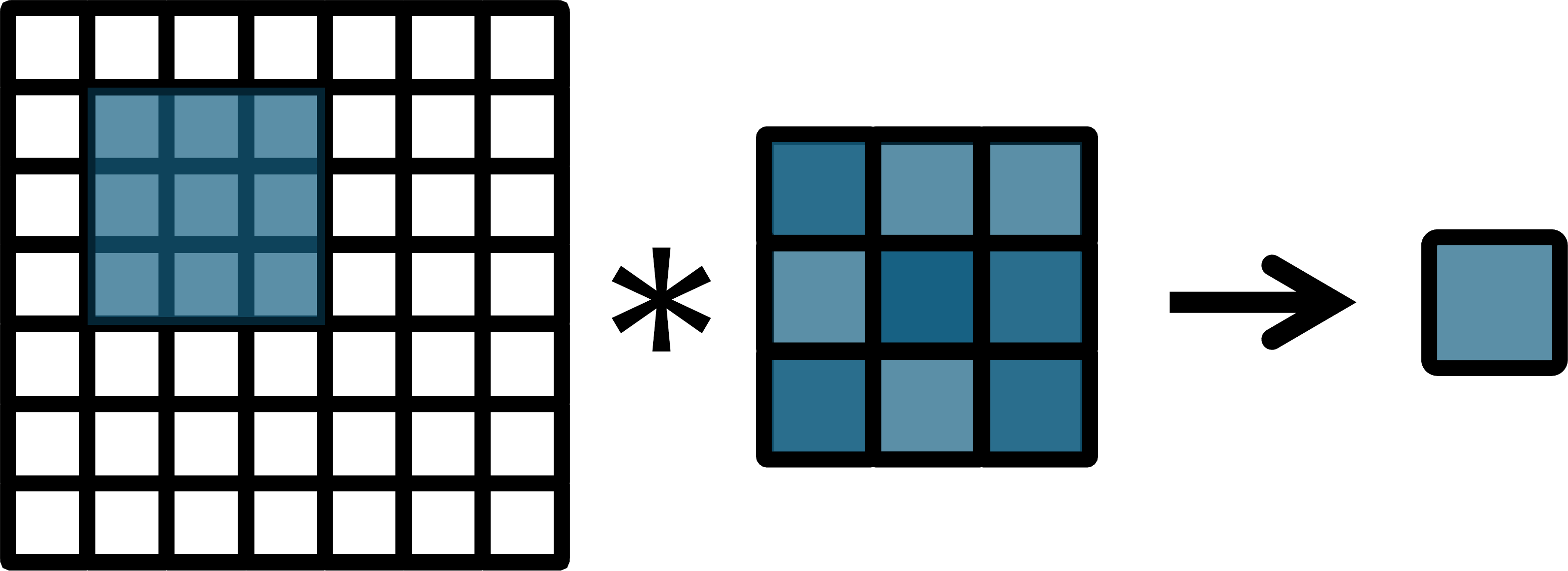} & \includegraphics[width=0.35\columnwidth]{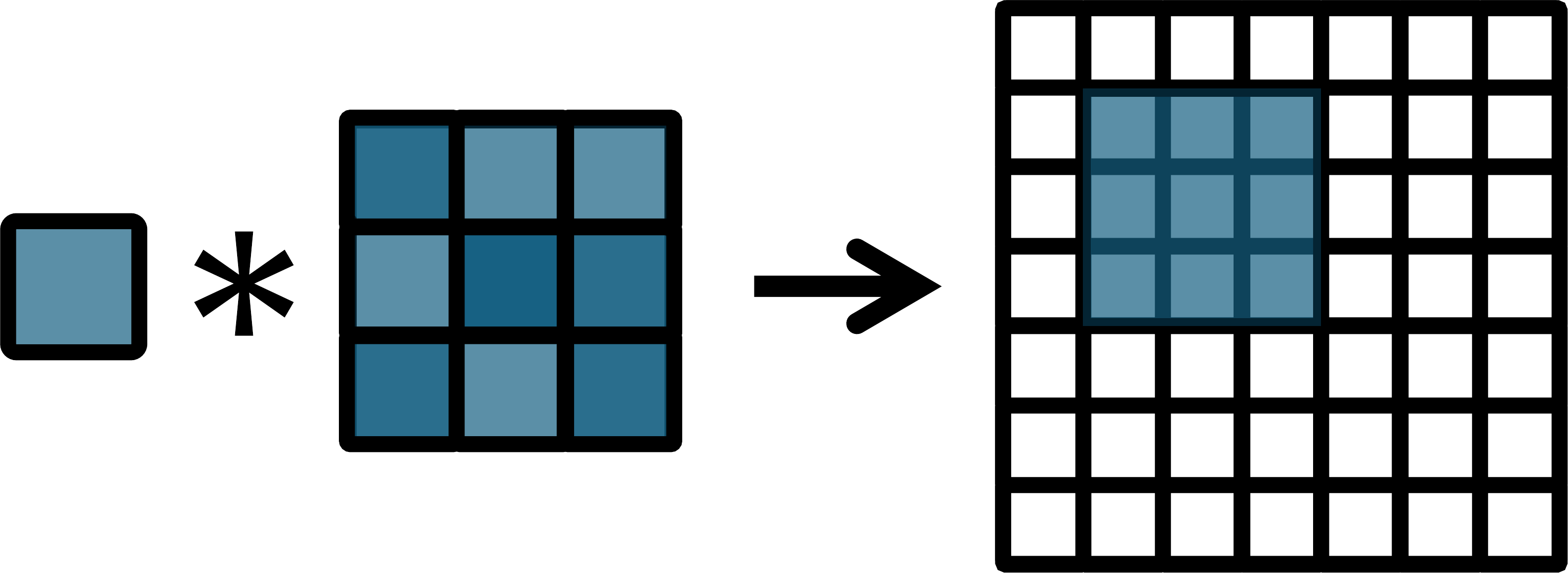}
    \end{tabular}
    \vspace{-0.3cm}
    \caption{
\textbf{Gathering versus scattering kernels.} Gathering kernels (a) weigh the neighbouring sparse samples for every output pixel. Sparsity is distributed \textit{within} the kernels, ensuring an even distribution of difficulty for the network's prediction task. Scattering kernels (b) scatter the sparse samples to neighbouring output pixels. Sparsity is distributed \textit{across} the kernels; kernels at sampled pixels are dense and solely responsible for the output image, while the majority of kernels at non-sampled pixels have no effect.}
    \label{fig:gather_scatter_concept}
\end{figure}

Many recent neural denoisers rely on kernel prediction rather than direct image synthesis. The predicted kernels define how noisy input samples are filtered to produce the final denoised image. As illustrated in \cref{fig:gather_scatter_concept}, kernel-based denoising can be formulated in two main ways: scattering, where each noisy pixel distributes its contribution to neighboring output pixels, and gathering, where each output pixel aggregates contributions from neighboring noisy input pixels.

With scattering, the network predicts a kernel at each sampled pixel that spreads its value to the surrounding area. When samples are dense, kernels overlap, and every output pixel receives contributions from multiple sources. However, as sampling becomes sparse, the network must predict meaningful kernels only at the sparse set of sampled locations, while the majority of kernel predictions go unused. This creates a complex task for the network: it must learn drastically different behavior depending on whether a pixel is sampled or not. %

Gathering transposes this relationship. The network predicts a kernel at each output pixel that reaches into the neighborhood to collect available samples. Even when samples are sparse, every output location has an associated kernel that controls how nearby information is integrated. The prediction task is uniformly distributed across all pixels, and the network learns a consistent strategy for interpolating sparse data. This formulation remains more stable when the average sampling rate drops below one sample per pixel, making it well-suited to the sparsity targeted in our work.

We adopt a multi-scale pyramid structure to provide a sufficient receptive field for reconstruction. Prior work~\citep{Balint2023Neural} constructs the pyramid through partitioning, a scatter-based operation, and uses $4 \times 4$ scatter kernels for coarse-to-fine upsampling. We instead build the pyramid via plain downsampling and use $2 \times 2$ gather kernels for upsampling, maintaining consistency with our gathering formulation throughout. At each of the five pyramid levels, the denoiser predicts spatially-varying $5 \times 5$ gather kernels for denoising. Reconstruction proceeds coarse-to-fine. At full resolution, we additionally predict $5 \times 5$ temporal kernels that gather from the warped previous HDR output. Full equations and implementation details are provided in \cref{sec:appendix_pyramid}.

\subsection{Learned Demodulation}
\label{sec:demodulation}

\begin{figure*}
    \centering
    \includegraphics[width=\textwidth]{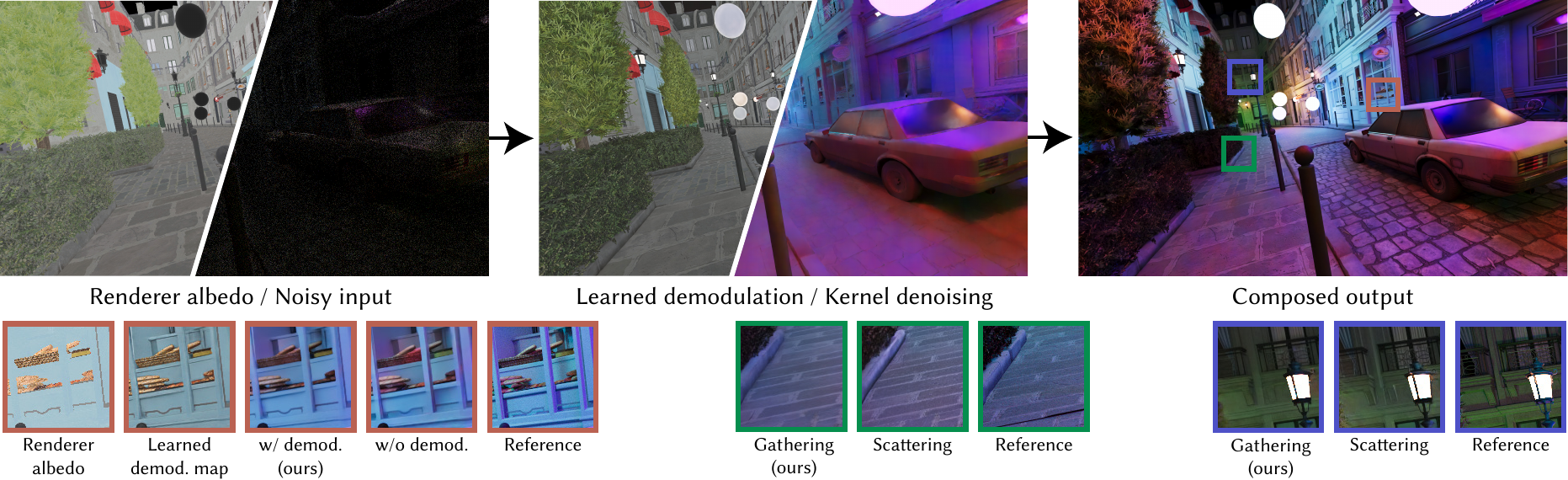}
    \vspace{-0.5cm}
    \caption{\textbf{Learned demodulation and effect of gather vs scatter pyramids.} Our sampler network receives the renderer's albedo buffer and the sparse, noisy input. Rather than requiring a mathematically correct albedo decomposition, our method learns a per-pixel demodulation map that adapts to the scene content. The learned map also encodes geometric shading cues resembling ambient occlusion, which are difficult to reconstruct through kernel filtering alone.
    \textit{Bottom left:} Using learned demodulation (w/ demod.) better preserves texture saturation and helps color bleeding compared to only kernel-based denoising (w/o demod.).
    \textit{Bottom middle and right:} Our gather-based pyramid produces cleaner reconstructions than the scatter-based alternative, which exhibits ringing and oversharpening artifacts around some structures.}
    \label{fig:albedo}
\end{figure*}

Kernel-based denoisers produce outputs that are weighted averages of neighboring input pixels. This constrains the output color at each pixel to lie within the convex hull of its neighbors' input colors. For illumination, this is not problematic, but high-frequency textures with saturated colors often require values outside this hull.

Prior albedo demodulation addresses this by dividing out the material's reflectance before filtering, then remodulating afterward. The denoiser operates on the illumination component, while texture detail is preserved through the albedo. However, this approach requires the renderer to provide mathematically correct demodulation maps for diffuse and specular components. Such decomposition is well-defined for simple materials but becomes problematic for volumes, measured BRDFs, and neural materials. It also fails when the albedo buffer is noisy or aliased, or when effects like motion blur and depth-of-field blend multiple surface points into a single pixel.

We sidestep these issues by letting the network learn its own per-pixel demodulation. Rather than requiring a precise albedo from the renderer, we provide a rich texture buffer that combines diffuse color, specular tint, and Fresnel information. The denoiser predicts a three-channel demodulation map and uses it to demodulate the noisy input before filtering. After reconstruction, we remodulate using the predicted map. This learned decomposition adapts to arbitrary materials without renderer-side changes and can also handle imperfect GBuffer inputs.

Kernel filtering is fundamentally a neighborhood operation where each output pixel is derived from its surrounding input pixels. The demodulation map also provides per-pixel control: the brightness adjustment at each location depends only on the predicted map value at that location, not on neighboring pixels. We observe that the network learns to encode some geometric shading into this map, producing darkening at corners and crevices that visually resembles ambient occlusion (\cref{fig:albedo}). Seemingly, such sharp, geometry-dependent variations could be more easily expressed as per-pixel multipliers predicted, rather than through the blending of noisy neighbors with intricately shaped kernels.

\subsection{Architecture and Training}
\label{sec:arch_training}

Both the sampler and denoiser are implemented as U-Net encoder-decoders. Following \citet{Salehi2022DeepAdaptive}, we adopt a global summary module at the sampler's bottleneck to aggregate image-wide statistics. We chose to use the 15M parameter architecture proposed by~\citet{Balint2023Neural}, a version of which also serves as the highest-quality preset for OIDN~\citep{OpenImageDenoise}. This well-tested architecture lets us evaluate the potential of our contributions under ideal conditions, keeping network limitations out of scope. Nonetheless, we show performance with a real-time network~\citep{Thomas2022Temporally} and discuss inference performance in \cref{sec:inference}.

We train a single model covering the entire sampling range from 0.11~spp to 4~spp. The discretization of the sample density map described in \cref{sec:sampling} is implemented using blue noise dithering \citep{Ulichney1993VoidandclusterMF}. We used the Falcor \citep{Kallweit22} research renderer and pre-rendered samples up to 256 spp to allow flexible budget selection during training. We extend the Noisebase dataset~\citep{Balint2023Neural} with these pre-rendered samples; the dataset contains 1024 training sequences of 64 frames each. Training runs for 256 epochs on a single H100 GPU, taking approximately 5 days. We use a batch size of 8 and train with 16-bit mixed-precision using the AdamW optimizer \citep{loshchilov2019decoupledweightdecayregularization} with $\beta = (0.8, 0.985)$ and a weight decay of 0.02. We decay the learning rate following a cosine annealing schedule \citep{loshchilov2017sgdrstochasticgradientdescent}.

We train on a sliding window of two consecutive frames to capture temporal dynamics. The loss operates in tonemapped sRGB space, combining a spatial term with a motion-compensated temporal term:
\begin{equation}
\label{eq:loss_spatial}
\mathcal{L}_{\text{spatial}} = \lvert \tdenoised - I^{\text{ref}} \rvert \cdot m\,,
\end{equation}
where $m$ is a visual masking term generated by the MILO perceptual metric~\citep{cogalan2025milo}, emphasizing visually more prominent absolute errors. The temporal term, following prior works \citep{hangming2025streaming, Thomas2022Temporally, Balint2023Neural}, penalizes frame-to-frame inconsistency:
\begin{equation}
\label{eq:loss_temporal}
\mathcal{L}_{\text{temporal}} = \lvert \Delta\tdenoised - \Delta I^{\text{ref}} \rvert\,,
\end{equation}
where $\Delta$ denotes the difference between the current frame and the warped previous frame. The final loss takes the maximum of both terms:
\begin{equation}
\label{eq:loss_combined}
\mathcal{L} = \max\bigl(1.25 \cdot \mathcal{L}_{\text{temporal}},\; \mathcal{L}_{\text{spatial}}\bigr)\,.
\end{equation}
We expect such temporal losses to be superseded by video generalizations of metrics like MILO in the near future, as we discuss in \cref{sec:temporal_stability}.

\begin{figure*}
    \centering
    \includegraphics[width=\textwidth]{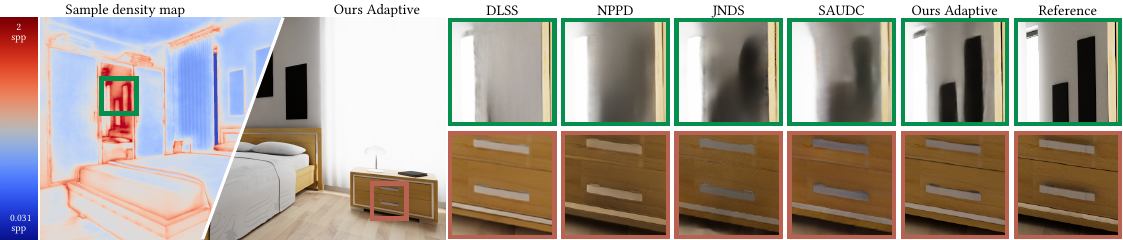}
    \includegraphics[width=\textwidth]{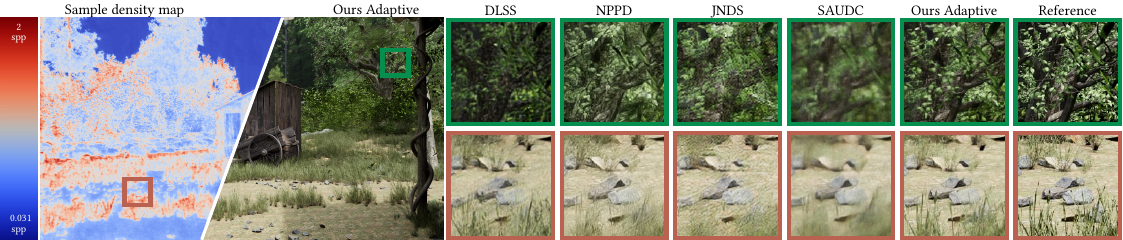}
    \vspace{-0.7cm}
    \caption{\textbf{Main comparison at 0.25~spp.} We compare our method against baselines at a 50\% superresolution equivalent budget. Our adaptive method preserves sharp edges and fine texture that superresolution approaches (DLSS, JNDS) lose due to low-resolution rendering. Other baselines also blur details more than our method. The density maps show our learned sample allocation: blue indicates fewer samples than uniform, red indicates more.}
    \label{fig:eval_main}
\end{figure*}

\section{Evaluation}
\label{sec:evaluation}

We evaluate our method across five aspects: comparison with prior work, budget scaling behavior, resolution scaling, the advantages of sparse sampling over superresolution, and component ablations. Our experiments demonstrate that adaptive sparse sampling enables high-quality denoising at budgets where previous methods fail, and each of our contributions provides improvements. We also analyze the robustness of adaptive sampling, showing that it reliably improves perceptually important regions without introducing failures. We also refer to the supplementary video for temporal qualitative assessment.

\paragraph{Setup.}
We evaluate across a range of rendering budgets from 0.11~spp (equivalent to 33\% superresolution scaling) to 4~spp. Our test set comprises 9 animated scenes with 160--300 frames each, rendered at 1440p unless otherwise noted. Reference images are rendered at 6144~spp.

\paragraph{Baselines.}
We compare against methods spanning three categories: superresolution, uniform denoising, and adaptive sampling.

\textit{DLSS 4 Ray Reconstruction}~\citep{nvidia_dlss4_research} denoises and upsamples from low-resolution renderings with low-resolution GBuffers. It supports variable scaling factors, making it suitable for budget comparisons. We use the official implementation in RTXPT~\citep{RTXPT} adapted to Falcor.

\textit{JNDS}~\citep{Thomas2022Temporally} similarly denoises and upsamples from low resolution but only supports 50\% superresolution scaling. As no official implementation is available, we implement the method in our framework, following the author's guidance and using a single filter path for combined diffuse and specular lighting, as they recommended.

\textit{NPPD}~\citep{Balint2023Neural} is a denoiser without adaptive sampling. We use the official implementation, retrained on sparse renderings with high-resolution GBuffers to match the inputs of our uniform version.

\textit{NTASD}~\citep{hasselgren2020ntasd} combines a denoiser with a learned sampler. We implement this method ourselves, as no official code is available, and reproduce the demonstrated performance at 4~spp. Training becomes unstable at lower budgets: at 2~spp, training diverged after 15--20 epochs across 10 attempts, from which we took the best-performing checkpoint. At 1~spp, training failed to complete even a single epoch.

\textit{SAUDC}~\citep{hangming2025streaming} extends adaptive sampling with mandatory samples for improved stability. Starting from the authors' official implementation, we found that training still diverged after 30--40 epochs at 2~spp despite mandatory samples. We replaced deterministic rounding with our stochastic formulation, which allowed training across our full budget range, though runs still diverged between epochs 80--100. We report results from the best checkpoint at epoch 83.

\textit{RLSNAS}~\citep{scardigli2023rl} uses reinforcement learning for sample allocation. The official implementation is designed for scene-specific training on simple scenes. We trained it on our full dataset until the RL framework's convergence criterion was met under the official configuration, but generalization to our test scenes remains poor.

Our method, across hundreds of training runs during experimentation, did not diverge once, except for some temporal-loop and gradient-estimator ablations noted below.

\paragraph{Metrics.}
We report six complementary metrics. PSNR measures pixel-wise error. MS-SSIM~\citep{wang2004image} and HaarPSI~\citep{reisenhofer2018haar} assess structural similarity at multiple scales. MILO~\citep{cogalan2025milo} predicts perceptual quality based on human opinion scores. CVVDP~\citep{mantiuk2024cvvdp} models visual differences in videos, accounting for contrast sensitivity and masking. CGVQM~\citep{jindal2025cgvqm} evaluates video quality across learned features, including temporal aspects. Higher values indicate better quality for all metrics.

\subsection{Main Comparison}
\label{sec:eval_main}

We first compare in the 50\% superresolution setting, the most common target in prior work. This corresponds to 0.25~spp sparse sampling at native resolution, or equivalently, upscaling from 720p to 1440p for superresolution methods. \Cref{tab:superres_sota} presents results across all metrics.

\begin{table}[ht]
    \centering
    \caption{\textbf{Comparison to baseline methods.} 50\% superresolution equivalent (0.25 spp sparse). Our method outperforms all baselines across all metrics. Higher is better for all metrics.}
    \label{tab:superres_sota}
    \resizebox{\columnwidth}{!}{%
    \begin{tabular}{lcccccc}
    \toprule
    \textbf{Method} & \textbf{PSNR}$\uparrow$ & \textbf{MS-SSIM}$\uparrow$ & \textbf{HaarPSI}$\uparrow$ & \textbf{MILO}$\uparrow$ & \textbf{CVVDP}$\uparrow$ & \textbf{CGVQM}$\uparrow$ \\
    \midrule
    \multicolumn{7}{l}{\tsubheader{Superresolution w/ LR inputs}} \\
    DLSS & 22.88 & 0.8950 & 0.6195 & 2.506 & 6.349 & 45.01 \\
    JNDS & 21.92 & 0.8657 & 0.5370 & 2.276 & 5.621 & 28.87 \\
    \multicolumn{7}{l}{\tsubheader{Sparse uniform sampling w/ HR inputs}} \\
    NPPD & 23.92 & 0.9062 & 0.6139 & 2.534 & 6.393 & 47.02 \\
    Ours Uniform & 24.64 & 0.9183 & 0.6532 & 2.640 & 6.753 & 55.21 \\
    \multicolumn{7}{l}{\tsubheader{Sparse adaptive sampling w/ HR inputs}} \\
    RLSNAS & 18.33 & 0.8077 & 0.3969 & 1.986 & 3.906 & -9.08 \\
    SAUDC & 23.09 & 0.8848 & 0.5645 & 2.250 & 5.965 & 38.97 \\
    Ours Adaptive & \textbf{25.41} & \textbf{0.9273} & \textbf{0.6815} & \textbf{2.677} & \textbf{7.056} & \textbf{57.63} \\
    \bottomrule
    \end{tabular}%
    }
\end{table}

Our method consistently outperforms all baselines by a substantial margin. Among superresolution approaches, DLSS achieves 22.88~dB PSNR while JNDS reaches only 21.92~dB, both limited by the loss of high-frequency detail inherent to low-resolution rendering. Sparse uniform sampling with high-resolution GBuffers outperforms superresolution: NPPD achieves 23.92~dB, while our uniform baseline reaches 24.64~dB. Adaptive sampling provides further gains, with our full method achieving 25.41~dB PSNR and 7.056~JOD CVVDP.

Visual comparisons in \cref{fig:eval_main} reveal that our method recovers fine details that competitors either blur or reconstruct with visible artifacts. The density maps show that our learned allocation strategy concentrates samples in such error-prone regions.

\begin{figure*}
    \centering
    \includegraphics[width=\textwidth]{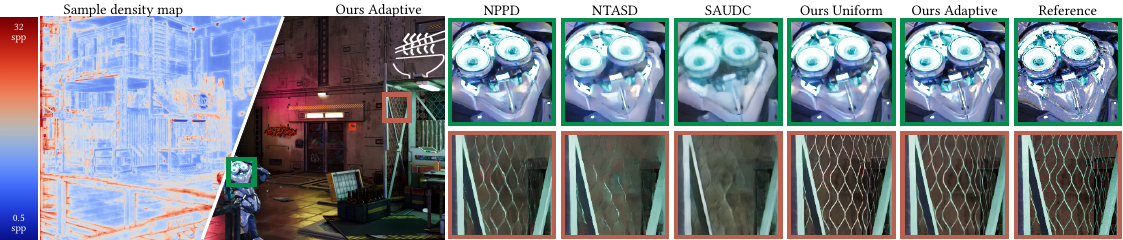}
    \includegraphics[width=\textwidth]{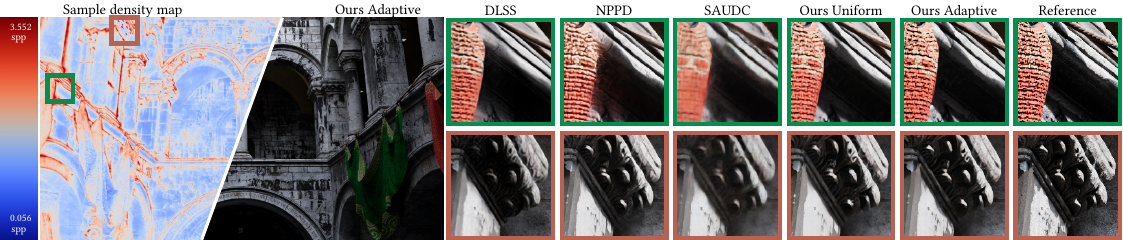}
    \includegraphics[width=\textwidth]{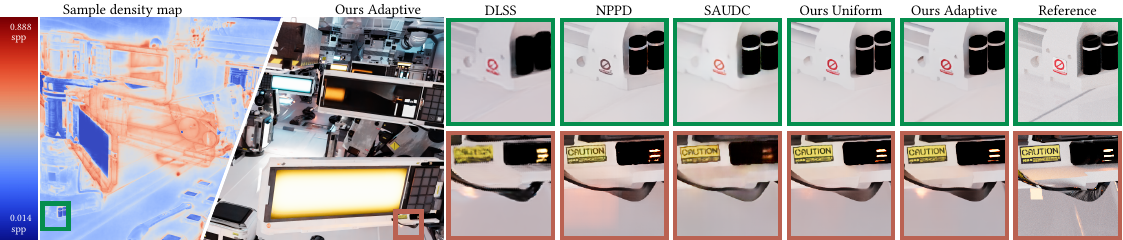}
    \vspace{-0.7cm}
    \caption{\textbf{Visual comparison across budgets and scenes.} We compare our method against baselines at 4.0~spp (top), 0.444~spp (middle), and 0.111~spp (bottom). Our adaptive method recovers fine details that competitors blur. The sample density maps show our learned sample allocation: blue indicates fewer samples than uniform, red indicates more. Our sampler concentrates samples on perceptually important regions such as edges and specular reflections while reducing allocation in smooth areas. }
    \label{fig:Main_figure_5}
\end{figure*}

\subsection{Budget Scaling}
\label{sec:eval_budget}

We evaluate scaling across two regimes: sub-1-spp (\cref{tab:sub1spp_scaling}) and above-1-spp (\cref{tab:sup1spp_scaling}). Budget levels correspond to DLSS presets for consistent comparison.

\begin{table}[ht]
    \centering
    \caption{\textbf{Sub-1-spp regime comparison.} Superresolution equivalent shown in parentheses. We show how our method and baselines scale as we lower the rendering budget. (PSNR$\uparrow$ / CVVDP$\uparrow$)}
    \label{tab:sub1spp_scaling}
    \resizebox{\columnwidth}{!}{%
    \begin{tabular}{lcccc}
    \toprule
     & \textbf{0.44 spp} & \textbf{0.34 spp} & \textbf{0.25 spp} & \textbf{0.11 spp} \\
    \textbf{Method} & (67\%) & (58\%) & (50\%) & (33\%) \\
    \midrule
    DLSS & 23.50/6.64 & 23.19/6.50 & 22.88/6.35 & 21.74/5.76 \\
    NPPD & 24.57/6.76 & 24.26/6.59 & 23.92/6.39 & 22.96/5.85 \\
    Ours Uniform & 25.46/7.14 & 25.07/6.96 & 24.64/6.75 & 23.40/6.18 \\
    \midrule
    SAUDC & 23.54/6.23 & 23.33/6.10 & 23.09/5.97 & 22.39/5.56 \\
    Ours Adaptive & \textbf{26.15/7.39} & \textbf{25.80/7.23} & \textbf{25.41/7.06} & \textbf{24.26/6.56} \\
    \bottomrule
    \end{tabular}%
    }
\end{table}

In the sub-1-spp regime, our method maintains stable quality with graceful degradation down to 0.11~spp (one sample per nine pixels on average). At this extreme sparsity, we achieve 24.26~dB PSNR compared to 21.74~dB for DLSS and 22.39~dB for SAUDC.

\begin{table}[ht]
    \centering
    \caption{\textbf{Above-1-spp regime comparison.} We show how our method and baselines scale to larger budgets. (PSNR$\uparrow$ / CVVDP$\uparrow$)}
    \label{tab:sup1spp_scaling}
    \resizebox{0.9\columnwidth}{!}{%
    \begin{tabular}{lcccc}
    \toprule
    \textbf{Method} & \textbf{1 spp} & \textbf{2 spp} & \textbf{3 spp} & \textbf{4 spp} \\
    \midrule
    NPPD & 25.44/7.23 & 26.08/7.56 & 26.47/7.75 & 26.70/7.86 \\
    Ours Uniform & 26.41/7.57 & 27.09/7.89 & 27.45/8.06 & 27.69/8.17 \\
    \midrule
    NTASD & --- & 18.34/3.98 & 24.84/6.77 & 25.91/7.27 \\
    SAUDC & 24.09/6.57 & 24.45/6.81 & 24.61/6.93 & 24.71/7.00 \\
    Ours Adaptive & \textbf{27.10/7.80} & \textbf{27.82/8.11} & \textbf{28.23/8.28} & \textbf{28.51/8.39} \\
    \bottomrule
    \end{tabular}%
    }
\end{table}

Above 1~spp, adaptive allocation continues to provide consistent gains. At 4~spp, our method achieves 28.51~dB PSNR versus 27.69~dB for our uniform version and 26.70~dB for NPPD.

In addition, \cref{fig:Main_figure_5} presents visual comparisons at average sampling rates of $4$~spp, $0.444$~spp, and $0.111$~spp. Compared to other methods, both our uniform and adaptive variants consistently achieve higher visual fidelity across all budgets. The benefit of adaptive sampling is especially apparent in fine details. For instance, it more accurately reconstructs the grid colors in the top scene, preserves specular highlights in the first crop of that scene, and improves detail recovery in the second crop of the last scene.

\subsection{Resolution Scaling}
\label{sec:eval_resolution}

A key advantage of sparse sampling over superresolution is favorable scaling with target resolution. We evaluate this effect by rendering four of our test scenes at 1080p, 1440p, and 4K, comparing our method against DLSS at both 912k and 2M total samples (\cref{tab:resolution_scaling}).

\begin{table}[ht]
    \centering
    \caption{
        \textbf{Comparison across resolutions.} We compare scaling across 3 resolutions, 2 budgets, and 6 metrics. Our adaptive method consistently outperforms our uniform version, which in turn outperforms the baseline DLSS. Comparison across resolutions is also dependent on the metric's scaling, since the reference images have different resolutions.
    }
    \label{tab:resolution_scaling}
    \resizebox{\columnwidth}{!}{%
    \begin{tabular}{lcccccc}
    \toprule
    \textbf{Method} & \textbf{PSNR}$\uparrow$ & \textbf{MS-SSIM}$\uparrow$ & \textbf{HaarPSI}$\uparrow$ & \textbf{MILO}$\uparrow$ & \textbf{CVVDP}$\uparrow$ & \textbf{CGVQM}$\uparrow$ \\
    \midrule
    \multicolumn{7}{l}{\tsubheader{1080p 1~spp (2M samples)}} \\
    Ours Adaptive & 27.69 & 0.9565 & 0.7667 & 3.003 & 7.890 & 69.25 \\
    Ours Uniform & 27.00 & 0.9521 & 0.7427 & 2.901 & 7.665 & 67.28 \\
    DLSS & 24.56 & 0.9258 & 0.6972 & 2.767 & 6.909 & 60.10 \\
    \multicolumn{7}{l}{\tsubheader{1440p 0.44~spp (2M samples)}} \\
    Ours Adaptive & 26.65 & 0.9408 & 0.7135 & 2.781 & 7.648 & 63.81 \\
    Ours Uniform & 26.01 & 0.9340 & 0.6859 & 2.707 & 7.363 & 61.91 \\
    DLSS & 23.99 & 0.9154 & 0.6564 & 2.594 & 6.886 & 54.55 \\
    \multicolumn{7}{l}{\tsubheader{4K 0.25~spp (2M samples)}} \\
    Ours Adaptive & 26.48 & 0.9340 & 0.6861 & 2.717 & 7.787 & 63.08 \\
    Ours Uniform & 25.80 & 0.9251 & 0.6557 & 2.639 & 7.445 & 60.92 \\
    DLSS & 23.87 & 0.9120 & 0.6283 & 2.524 & 6.955 & 52.58 \\
    \multicolumn{7}{l}{\tsubheader{1080p 0.44~spp (912k samples)}} \\
    Ours Adaptive & 26.65 & 0.9451 & 0.7272 & 2.842 & 7.486 & 64.62 \\
    Ours Uniform & 25.96 & 0.9369 & 0.6995 & 2.754 & 7.187 & 62.16 \\
    DLSS & 23.72 & 0.9132 & 0.6544 & 2.594 & 6.567 & 52.83 \\
    \multicolumn{7}{l}{\tsubheader{1440p 0.25~spp (912k samples)}} \\
    Ours Adaptive & 25.84 & 0.9299 & 0.6817 & 2.676 & 7.315 & 60.09 \\
    Ours Uniform & 25.11 & 0.9178 & 0.6506 & 2.595 & 6.916 & 57.48 \\
    DLSS & 23.41 & 0.9056 & 0.6240 & 2.485 & 6.637 & 48.55 \\
    \multicolumn{7}{l}{\tsubheader{4K 0.11~spp (912k samples)}} \\
    Ours Adaptive & 25.20 & 0.9179 & 0.6402 & 2.580 & 7.285 & 57.61 \\
    Ours Uniform & 24.43 & 0.9016 & 0.6048 & 2.491 & 6.797 & 54.29 \\
    DLSS & 22.78 & 0.8927 & 0.5674 & 2.354 & 6.522 & 42.74 \\
    \bottomrule
    \end{tabular}%
    }
\end{table}

Adaptive sampling maintains its advantage over uniform sampling across all conditions, with both methods outperforming DLSS by a wide margin. The advantage of adaptive over uniform is approximately 0.7~dB in PSNR, stable across resolutions and budgets.

As resolution increases at a fixed budget, per-pixel metrics decrease, as expected, since each pixel receives fewer samples while the resolution of the reference images increases. This effect is more pronounced for our CNN-based networks than for DLSS, whose transformer architecture appears to scale more favorably with increasing resolution. However, the benefit of additional samples also grows with resolution: at 1080p, doubling the budget from 912k to 2M samples improves our adaptive method by 1.04~dB PSNR, while at 4K the same increase yields 1.28~dB. DLSS shows a similar trend (0.84~dB to 1.09~dB).

\subsection{Sparse Sampling versus Superresolution}
\label{sec:eval_sparse_vs_sr}

To understand why sparse sampling outperforms superresolution at equivalent budgets, we run specific ablations at 0.25~spp (\cref{tab:sparse_vs_superres}).

We first test our uniform method with low-resolution inputs bilinearly upscaled to full resolution, matching the information available to superresolution methods. This includes path-traced samples, normals, albedo, and motion vectors, all at reduced resolution. Quality decreases (from 24.64~dB to 24.16~dB in PSNR), yet remains state-of-the-art compared to all baselines. This suggests that our contributions beyond adaptive sampling---the gather pyramid, learned demodulation, and TMO-aware training---could benefit superresolution pipelines.%

\begin{table}[ht]
    \centering
    \caption{\textbf{Sparse sampling versus superresolution.} We isolate factors contributing to sparse sampling's advantage at 0.25~spp (50\% superresolution equivalent). High-resolution GBuffers recover most of the quality gap, while density map resolution has small impact on aggregate metrics.}
    \label{tab:sparse_vs_superres}
    \resizebox{\columnwidth}{!}{%
    \begin{tabular}{lcccccc}
    \toprule
    \textbf{Method}$\uparrow$ & \textbf{PSNR}$\uparrow$ & \textbf{MS-SSIM}$\uparrow$ & \textbf{HaarPSI}$\uparrow$ & \textbf{MILO}$\uparrow$ & \textbf{CVVDP}$\uparrow$ & \textbf{CGVQM}$\uparrow$ \\
    \midrule
    Ours Uniform & 24.64 & 0.9183 & 0.6532 & 2.640 & 6.753 & 55.21 \\
    w/ LR input & 24.16 & 0.9088 & 0.6304 & 2.498 & 6.559 & 50.52 \\
    w/ LR inp., HR GB. & 24.40 & 0.9140 & 0.6413 & 2.609 & 6.641 & 54.00 \\
    \midrule
    \midrule
    Ours Adaptive & \textbf{25.41} & \textbf{0.9273} & \textbf{0.6815} & 2.677 & \textbf{7.056} & \textbf{57.63} \\
    w/ LR density map & 25.32 & 0.9265 & 0.6772 & \textbf{2.681} & 7.033 & 56.90 \\
    \bottomrule
    \end{tabular}%
    }
\end{table}

Restoring high-resolution GBuffers while keeping low-resolution radiance samples and motion recovers much of the quality (24.40~dB PSNR). High-resolution normals, depth, and albedo provide valuable guidance for reconstruction even when the radiance signal is sparse, suggesting that rendering with high-resolution GBuffers is a practical way to boost sharpness with minimal overhead, as we show in \cref{sec:gbuf-speed}.

For adaptive sampling, we test the impact of density map resolution by downsampling predictions to quarter resolution before stochastic sampling. The effect on aggregate metrics is small, but visual inspection reveals more accurate sampling of specular highlights and contact shadows at full resolution. These features occupy a small fraction of pixels, explaining the minimal metric difference, though they may contribute disproportionately to perceived quality. Running the sampler network at reduced resolution remains a viable trade-off in terms of efficiency.

\subsection{Component Ablations}
\label{sec:eval_ablations}

We validate each proposed component through ablations (\cref{tab:ablations}), evaluating both our adaptive and uniform models to isolate contributions.

\begin{table}[ht]
    \centering
    \caption{\textbf{Component ablations.} We validate each proposed component by replacing or removing it from our adaptive and uniform models. Loss formulation and recurrent features have the largest impact; our gradient estimator provides the best stability-accuracy balance. (PSNR$\uparrow$ / CVVDP$\uparrow$)}
    \label{tab:ablations}
    \resizebox{\columnwidth}{!}{%
    \begin{tabular}{lcccc}
    \toprule
    \textbf{Method} & \textbf{0.11 spp} & \textbf{0.25 spp} & \textbf{1 spp} & \textbf{4 spp} \\
    \midrule
    Ours Adaptive & 24.26/6.56 & 25.41/7.06 & 27.10/7.80 & 28.51/8.39 \\
    \multicolumn{5}{l}{\tsubheader{Gradient variants}} \\
    DASR surrogate & 23.58/6.23 & 25.00/6.83 & 26.86/7.67 & 28.28/8.31 \\
    Straight-through & 24.32/6.51 & 25.40/7.01 & 27.02/7.76 & 28.41/8.37 \\
    Gumbel-Softmax & 24.09/6.42 & 25.32/6.99 & 27.10/7.80 & 28.56/8.41 \\
    \multicolumn{5}{l}{\tsubheader{Component replacements}} \\
    SMAPE loss & 23.25/6.18 & 24.57/6.71 & 26.46/7.51 & 28.02/8.16 \\
    Photographic TMO loss & 23.49/6.28 & 24.68/6.81 & 26.53/7.62 & 28.18/8.31 \\
    Partitioning pyramid & 23.68/6.38 & 25.15/6.98 & 27.13/7.81 & 28.76/8.49 \\
    w/o learned demod. & 24.33/6.59 & 25.53/7.10 & 27.20/7.84 & 28.62/8.44 \\
    Separate rec. features & 23.96/6.35 & 25.03/6.82 & 26.65/7.62 & 27.98/8.26 \\
    w/o rec. features & 23.47/6.10 & 24.63/6.59 & 26.30/7.41 & 27.72/8.11 \\
    \midrule
    \midrule
    Ours Uniform & 23.40/6.18 & 24.64/6.75 & 26.41/7.57 & 27.69/8.17 \\
    \multicolumn{5}{l}{\tsubheader{Uniform ablations}} \\
    SMAPE loss & 22.61/5.95 & 23.91/6.47 & 25.74/7.31 & 27.12/7.97 \\
    Photographic TMO loss & 23.38/6.16 & 24.68/6.74 & 26.45/7.55 & 27.81/8.19 \\
    Partitioning pyramid & 22.97/6.11 & 24.37/6.72 & 26.28/7.58 & 27.59/8.20 \\
    w/o learned demod. & 23.49/6.29 & 24.76/6.86 & 26.50/7.63 & 27.79/8.22 \\
    \bottomrule
    \end{tabular}%
    }
\end{table}

\paragraph{Gradient estimation.}
We compare four gradient estimation strategies for adaptive sampling. With the DASR surrogate~\citep{Kuznetsov2018Deep}, training was unstable with only one of five runs succeeding. The resulting quality is also worse due to the biased gradients that guide the optimization toward suboptimal sample distributions. Straight-through estimation also proved unstable (three of five runs succeeded), but surprisingly achieves comparable quality when it converges, suggesting that its gradient bias may act as a regularizer, as observed in other domains \citep{bengio2013estimating}. Gumbel-Softmax trains stably but suffers from train-test mismatch as discussed in \cref{sec:relaxed_stoch_rounding}, slightly degrading quality at low sample counts where the contribution scaling in \cref{eq:stoch_rnoisy} amplifies discrepancies. Our relaxed formulation achieves the best balance of stability and accuracy.

\paragraph{Loss formulation.}
Replacing our tonemapping-aware pipeline with the SMAPE-based loss from NPPD~\citep{Balint2023Neural} substantially degrades quality for both adaptive and uniform variants, demonstrating the importance of perceptual alignment for training. Using a fixed photographic TMO instead of our randomized parametric family also hurts the adaptive model, confirming that adaptive sampling adjusts to tonemapped appearance and requires diverse TMOs for generalization. The uniform model is less affected since it cannot control sampling or observe the tonemapped image.

\paragraph{Filter architecture.}
Replacing our gather pyramid with NPPD's scatter-based partitioning pyramid degrades quality, particularly at low sample counts, confirming our analysis in \cref{sec:kernels} that scatter-based approaches are poorly suited to sparse inputs.

\paragraph{Learned demodulation.}
Removing learned demodulation slightly improves aggregate metrics. However, this component is essential for resolving color artifacts in regions with saturated albedos, as shown in \cref{fig:albedo}. Deriving more robust formulations holds potential for future work.

\paragraph{Recurrent features.}
Separating the recurrent feature spaces for the sampler and denoiser degrades quality and increases training variance, suggesting the shared latent space enables beneficial collaboration between networks. Removing recurrent features entirely led to further degradation and needed reruns due to divergence, confirming the importance of temporal context for stability and performance.

\begin{figure*}
    \centering
    \includegraphics[width=\textwidth]{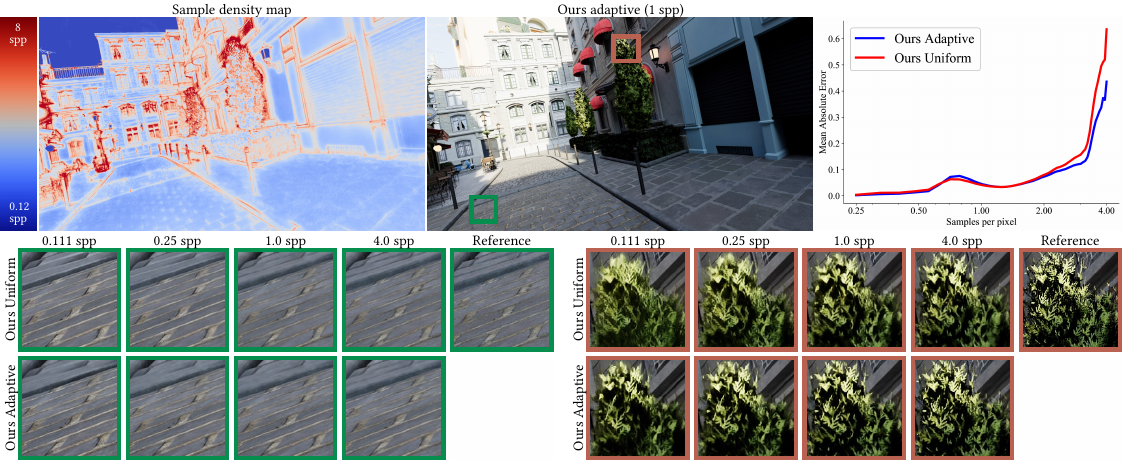}
    \vspace{-0.7cm}
    \caption{\textbf{Best-worst case robustness analysis.} This figure compares, for the same scene, the best- and worst-performing regions of our adaptive sampling method relative to uniform sampling. The green inset highlights the crop where uniform sampling performs best, while the orange inset highlights the crop where adaptive sampling performs best. The comparison shows that adaptive sampling better reconstructs fine details, while only slightly underperforming in more uniform regions. The plot on the right illustrates the mean absolute error as a function of relative sampling rate. When adaptive sampling allocates fewer samples than uniform sampling (left side of the plot, sampling ratio $< 1$), the error increases only marginally. In contrast, when adaptive sampling allocates more samples than uniform sampling (right side, sampling ratio $> 1$), the error reduction is substantial. This demonstrates that the sampler successfully focuses computation on high-error regions, leading to an overall improvement in image quality.}%
    \label{fig:eval_robustness}
\end{figure*}

\subsection{Robustness Analysis}
\label{sec:eval_robustness}

A fundamental concern with adaptive sampling is reliability: since the total sample budget is fixed, improving quality in some regions necessarily degrades it in others. Within our exploration-exploitation interpretation, the sampler must reliably identify which regions require more samples and which can afford fewer. We demonstrate that our method addresses regions with the most visible distortions while taking samples only from regions where the impact is perceptually negligible.

\Cref{fig:eval_robustness} presents a balanced comparison. For the given test frame, we identify the patch where adaptive sampling degrades quality the most compared to uniform sampling (measured by mean absolute error, MAE), as well as the patch where adaptive sampling improves quality the most. The figure shows both extremes: regions where adaptive sampling wins, such as fine geometric detail on tree leaves, and regions where it loses to uniform sampling, such as the pavement.

The key observation is that adaptive sampling wins where it matters. The patches showing improvement contain perceptually critical content, while the patches showing degradation are invariably smooth regions, such as diffuse surfaces, where error is difficult to perceive, as contrast is low.

The accompanying error-versus-density plots quantify this relationship. Our method predicts high density where reconstruction error would otherwise be large, thereby substantially reducing it. In regions predicted to require few samples, the error is consistently low as the sampler identifies simpler regions with high confidence. Adaptive sampling primarily loses to uniform sampling in intermediate regions where error is present but modest; here, the density map is only slightly below average, reflecting appropriate caution. We provide similar analyses for additional test scenes in \cref{sec:additional_results}.

\section{Discussion}

\subsection{Inference cost}
\label{sec:inference}

We have not investigated specific network architectures in depth, as inference performance is highly hardware-dependent and lightweight architectures remain a rapidly evolving field. Recent developments in low-precision quantized inference and knowledge distillation from larger to smaller models~\citep{Kong2025real} show considerable promise for enabling better real-time neural denoising. 

Unfortunately, state-of-the-art networks are not widely available or well-tested across methods. DLSS, for instance, employs a seemingly quite capable transformer architecture, but its architectural details are not public. Developing a custom real-time architecture was beyond the scope of this work.

Nonetheless, we recognize that some readers may question whether real-time networks will match the capacity of the 15M parameter architecture used in our main evaluation before full real-time path tracing becomes widespread. To address this concern, we also evaluate our method using the network from JNDS~\citep{Thomas2022Temporally}, a 2.6M-parameter architecture designed for real-time inference on current Intel GPUs.

\begin{table}[ht]
    \centering
    \caption{\textbf{Ablations using real-time network.} 50\% superresolution equivalent (0.25 spp sparse). Higher is better for all metrics.}
    \label{tab:network_scaling}
    \resizebox{\columnwidth}{!}{%
    \begin{tabular}{lcccccc}
    \toprule
    \textbf{Method} & \textbf{PSNR}$\uparrow$ & \textbf{MS-SSIM}$\uparrow$ & \textbf{HaarPSI}$\uparrow$ & \textbf{MILO}$\uparrow$ & \textbf{CVVDP}$\uparrow$ & \textbf{CGVQM}$\uparrow$ \\
    \midrule
    Ours Uniform & 24.64 & 0.9183 & 0.6532 & 2.640 & 6.753 & 55.21 \\
    Mini Uniform & 24.16 & 0.9050 & 0.6160 & 2.510 & 6.443 & 49.46 \\
    NPPD & 23.92 & 0.9062 & 0.6139 & 2.534 & 6.393 & 47.02 \\
    \midrule
    Ours Adaptive & 25.41 & 0.9273 & 0.6815 & 2.677 & 7.056 & 57.63 \\
    Mini Sampler & 25.37 & 0.9243 & 0.6791 & 2.735 & 6.961 & 57.70 \\
    Mini Adaptive & 24.97 & 0.9192 & 0.6528 & 2.637 & 6.820 & 54.30 \\
    \bottomrule
    \end{tabular}%
    }
\end{table}

As shown in \cref{tab:network_scaling}, our method scales well to reduced network capacity. The Mini Uniform variant, using the 2.6M parameter JNDS network, outperforms the strongest baseline NPPD on most metrics despite NPPD employing the same 15M parameter architecture as our main evaluation. Mini Sampler, which substitutes only the sampler network while retaining the large denoiser, incurs negligible performance loss. Mini Adaptive, with both a lightweight sampler and denoiser, still surpasses all baselines and performs comparably to our uniform model using the larger network. These results indicate that adaptive sampling continues to provide meaningful improvements even with architectures real-time-capable today.

\subsection{GBuffer \& Rendering time}
\label{sec:gbuf-speed}

Decreasing reliance on GBuffers is a promising direction of current research. For example, \citet{wu2024gffe} propose GBuffer-free extrapolation for high-framerate, low-latency rendering. Skipping rendering for the extrapolated frames reduces load on the game engine and saves time on preliminary tasks such as animation, BVH updates, and scheduling. However, scaling the GBuffer resolution only offers modest savings, as the GPU time spent on GBuffer rendering is relatively negligible to begin with. By comparison, tracing the scattered, incoherent rays for global illumination is inherently harder. 

In our setup, one spp uniform path tracing is generally 20-30 times the cost of GBuffer rendering at the same resolution. That is despite Falcor using a relatively unoptimized ray-traced GBuffer pass. In a production game engine, we expect GBuffer rendering to be much faster.

\begin{table}[ht]
\centering
\caption{\textbf{GBuffer timing comparison.} Cost of GBuffer and path tracing on several of our production quality test scenes, measured on a workstation with an NVIDIA RTX3090 at 1440p. PT adaptive does not include inference of the sample density.}
\label{tab:gbuffer}
\resizebox{\columnwidth}{!}{%
\begin{tabular}{llllll}
\toprule
\textbf{Scene} & \textbf{Triangles} & \textbf{GBuffer } & \textbf{PT uniform} & \textbf{PT 1/4 spp} & Budget \\
 &   & \textbf{(ms)} & \textbf{(ms)} & \textbf{adaptive (ms)} &  \\
\midrule
Bistro & 4.2M & 0.672 & 21.80 & 5.5 & $\approx$ 27\%\\
Zero-Day & 5.3M & 0.892 & 14.42 & 3.6 & $\approx$ 29\%\\
Sun Temple & 600k & 0.41 & 8.47 & 2.11 & $\approx$ 28\%\\
\bottomrule
\end{tabular}%
}
\end{table}

Time-wise scaling is further imperfect, as adaptive sampling tends to focus on more challenging parts of the scene, for example, prioritising sampling areas with complex lighting and geometry over flat regions. These regions sometimes slow down path tracing due to increased BVH and material complexity.

Nonetheless, as shown in \cref{tab:gbuffer}, the overhead of high-resolution GBuffers and adaptive sampling is marginal compared to the improvement over superresolution demonstrated in our work. We focus on sampling budget comparisons for the majority of our paper, as they are more practical and easier to follow. One may consider a $\approx 5\%$ overhead on our methods when translating sample count measurements to rendering time.

\subsection{Temporal stability}
\label{sec:temporal_stability}

Our sample density maps reveal that the sampler network compensates heavily for specular reflections, where motion vectors derived from primary surface intersections are unreliable. 

Furthermore, a significant task of the sampler network is detecting disocclusions, which improved reprojection methods could address more directly. With better temporal information, the denoiser could potentially predict sampling requirements for the next frame, allowing a much lighter sampler network to make only minor corrections.

Better motion vector estimation~\citep{Zeng2021temporally} and novel reprojection methods such as GBuffer-free frame extrapolation~\citep{wu2024gffe} are therefore promising directions for future work.

\section{Conclusion}

We presented the first end-to-end adaptive sampling and denoising pipeline designed for the sub-1-spp regime. Our stochastic formulation for sample placement enables stable gradient estimation at extreme sparsity, where prior deterministic approaches fail. Combined with tonemapping-aware training using perceptual losses and a gather-based pyramidal filter suited to sparse inputs, our method consistently outperforms both superresolution and previous adaptive sampling approaches across all tested budgets.

Our results demonstrate that adaptive sampling remains effective even when the average budget falls below one sample per pixel. The learned sampler reliably concentrates samples on perceptually critical regions while safely reducing allocation in smooth areas where errors are difficult to perceive.

\begin{acks}
We thank Anton Kaplanyan, Anton Sochenov, and Attila \'Afra for insightful discussions.

We gratefully acknowledge the support of the Saarland/Intel Joint Program on the Future of Graphics and Media. 

We thank the following authors for the scenes used in our experiments: Evermotion (Training scenes), Leartes Studios (Cyberpunk Rooftop Market, Village House), \citet{sponza22} (Sponza), \citet{ORCAAmazonBistro} and Krzysztof Wolski (Bistro), SlykDrako and \citet{resources16} (Bedroom), \citet{ZeroDay} (ZeroDay).
\end{acks}

\bibliographystyle{ACM-Reference-Format}
\bibliography{paper}

\newpage
\appendix

\section{Gathering pyramid definition}
\label{sec:appendix_pyramid}

Our pyramidal filter builds on NPPD~\citep{Balint2023Neural} but replaces scatter-based operations with gather-based equivalents. With scatter kernels, the network must predict meaningful weights only at sampled pixels while most predictions go unused; with gather kernels, every output pixel predicts weights to collect from its neighborhood, distributing the task uniformly regardless of input sparsity.

We also simplify the pyramid by removing learned partitioning. NPPD predicts per-pixel weights that partition radiance across levels before downsampling. We instead use plain average pooling and rely on joint softmax normalization across kernel types to implicitly balance contributions.

\paragraph{Pyramid construction.}
We build a low-pass pyramid with $L=5$ levels by average pooling:
\begin{equation}
\label{eq:app-pyramid}
\rlayernoisy^l_{\xy t} = \frac{1}{4^{l}} \sum_{u,v \in \mathcal{N}^l_{xy}} \rnoisy_{uvt} \text{,}
\end{equation}
where $\mathcal{N}^l_{xy}$ denotes the $2^l \times 2^l$ block of full-resolution pixels mapping to coarse pixel $(x,y)$ at level $l$. Level $l=0$ is full resolution; $l=L-1$ is coarsest.

\paragraph{Kernel prediction.}
At each level, we predict spatially varying gather kernels:
\begin{itemize}[nosep]
    \item $5 \times 5$ denoising kernels $\kerneldenoise^l$ at all levels (25 weights),
    \item $2 \times 2$ upsampling kernels $\kernelups^l$ at levels $l < L-1$ (4 weights),
    \item $5 \times 5$ temporal kernels $\kerneltemporal$ at $l = 0$ only (25 weights).
\end{itemize}
We apply joint softmax normalization across all kernels at each level. At the coarsest level, only denoising weights are normalized. At intermediate levels:
\begin{equation}
\label{eq:app-softmax-mid}
\sum_{ij} \kerneldenoise^{l}_{ij, \xy t} + \sum_{ij} \kernelups^{l}_{ij, \xy t} = 1 \text{.}
\end{equation}
At full resolution, the softmax spans all 54 weights:
\begin{equation}
\label{eq:app-softmax-full}
\sum_{ij} \kerneldenoise^{0}_{ij, \xy t} + \sum_{ij} \kernelups^{0}_{ij, \xy t} + \sum_{ij} \kerneltemporal_{ij, \xy t} = 1 \text{.}
\end{equation}
This joint normalization eliminates separate blending coefficients: the network learns to balance denoising, upsampling, and temporal contributions through relative kernel magnitudes.

\paragraph{Gather filtering.}
Each output pixel collects from its $5 \times 5$ neighborhood:
\begin{equation}
\label{eq:app-gather}
G(\rlayernoisy^l, \kerneldenoise^l)_{\xy t} = \sum_{i,j=-2}^{2} \kerneldenoise^l_{ij, \xy t} \cdot \rlayernoisy^l_{(x+i)(y+j)t} \text{.}
\end{equation}

\paragraph{Learnable bilinear upsampling.}
Standard bilinear interpolation computes each output pixel as a fixed weighted average of four neighbors. We replace the fixed weights with learned per-pixel weights, allowing edge-aware upsampling. For upsampling from level $l+1$ to level $l$:
\begin{equation}
\label{eq:app-upsample}
U(\rlayerdenoised^{l+1}, \kernelups^l)_{\xy t} = \sum_{i,j=0}^{1} \kernelups^l_{ij, \xy t} \cdot \rlayerdenoised^{l+1}_{\lfloor \frac{x+i}{2} \rfloor \lfloor \frac{y+j}{2} \rfloor t} \text{.}
\end{equation}
The four kernel weights replace the position-dependent bilinear coefficients, enabling the network to sharpen edges or smooth across them as needed.

\paragraph{Coarse-to-fine reconstruction.}
Starting from the coarsest level:
\begin{align}
\label{eq:app-recon-coarse}
\rlayerdenoised^{L-1}_{\xy t} &= G(\rlayernoisy^{L-1}, \kerneldenoise^{L-1})_{\xy t} \text{,} \\[2mm]
\label{eq:app-recon-mid}
\rlayerdenoised^{l}_{\xy t} &= G(\rlayernoisy^{l}, \kerneldenoise^{l})_{\xy t} + U(\rlayerdenoised^{l+1}, \kernelups^{l})_{\xy t} \text{,}
\end{align}
for $l = L-2, \ldots, 1$. At full resolution, we add temporal filtering:
\begin{multline}
\label{eq:app-recon-full}
\rdenoised_{\xy t} = G(\rnoisy, \kerneldenoise^{0})_{\xy t} + U(\rlayerdenoised^{1}, \kernelups^{0})_{\xy t} \\
+ G(\mathcal{W}_t \rdenoised_{t-1}, \kerneltemporal)_{\xy t} \text{,}
\end{multline}
where $\mathcal{W}_t$ warps the previous frame using motion vectors. The joint softmax ensures that in disoccluded regions or with unreliable motion, the network suppresses temporal contribution by reducing $\kerneltemporal$ weights.

\section{Additional results}
\label{sec:additional_results}

\begin{figure*}
    \centering
    \includegraphics[width=0.9\textwidth]{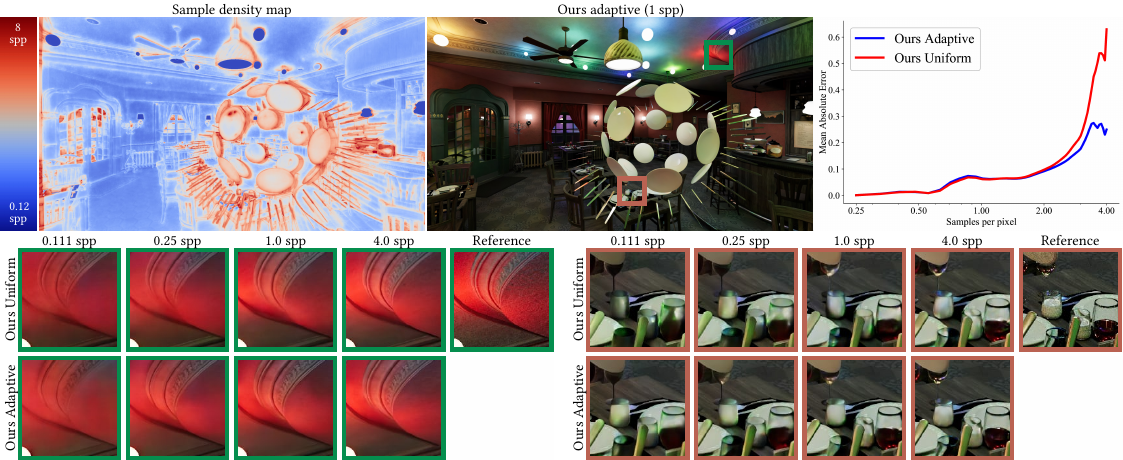}
    \includegraphics[width=0.9\textwidth]{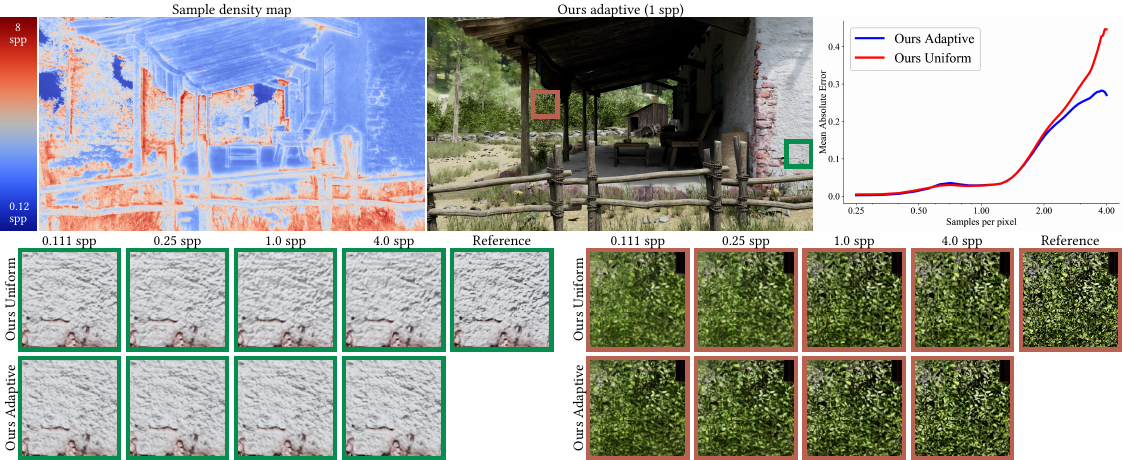}
    \includegraphics[width=0.9\textwidth]{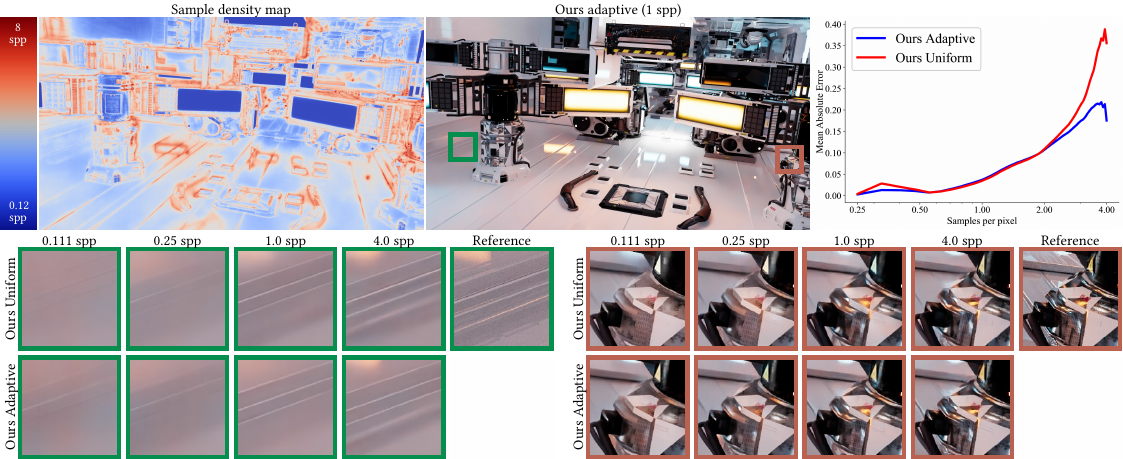}

    \caption{\textbf{Best-worst case robustness analysis.} This figure presents additional results analogous to \Cref{fig:eval_robustness} on different scenes. It highlights the best- and worst-performing regions of our adaptive method compared to the uniform baseline. The plot on the right shows the mean absolute error as a function of the relative sampling rate: regions undersampled by our method appear on the left, while oversampled regions appear on the right.}
    \label{fig:eval_robustness_many}
\end{figure*}

\Cref{fig:eval_robustness_many} presents additional comparisons extending \cref{fig:eval_robustness} to three further scenes. For each scene, we show the best- and worst-performing crops of our adaptive method relative to the uniform baseline. These results demonstrate that the adaptive sampler consistently reallocates the sampling budget toward visually complex regions with fine details, leading to noticeable improvements in quality, while preserving high reconstruction quality in simpler regions.\\

In addition to these figures, we provide an interactive image viewer in the supplement to facilitate detailed visual comparisons between methods, as well as a video that further illustrates temporal behavior and perceived quality.

\end{document}